\documentclass[journal,twoside,web]{ieeecolor}

\usepackage{etoolbox}
\makeatletter
\@ifundefined{color@begingroup}%
  {\let\color@begingroup\relax
   \let\color@endgroup\relax}{}%
\def\fix@ieeecolor@hbox#1{%
  \hbox{\color@begingroup#1\color@endgroup}}
\patchcmd\@makecaption{\hbox}{\fix@ieeecolor@hbox}{}{\FAILED}
\patchcmd\@makecaption{\hbox}{\fix@ieeecolor@hbox}{}{\FAILED}

\usepackage{tmi}
\usepackage{cite}
\usepackage{amsmath,amssymb,amsfonts}
\usepackage{algorithmic}
\usepackage{graphicx}
\usepackage{textcomp}

\usepackage{multirow}
\usepackage{booktabs}
\usepackage{pifont}
\usepackage{hyperref}
\usepackage{makecell}

\def\BibTeX{{\rm B\kern-.05em{\sc i\kern-.025em b}\kern-.08em
    T\kern-.1667em\lower.7ex\hbox{E}\kern-.125emX}}
\begin{document}

\title{DeepSparse: A Foundation Model for Sparse-View CBCT Reconstruction}

\author{
Yiqun Lin$^{\dag}$, 
Jixiang Chen$^{\dag}$, 
Hualiang Wang, 
Jiewen Yang,
Jiarong Guo,
Yi Zhang,
Xiaomeng Li
\thanks{$^{\dag}$ indicates equal contribution.}
\thanks{
This work is partially supported by a research grant from the National Natural Science Foundation of China under Grant 62306254 and a grant from the Hong Kong Innovation and Technology Fund under Grant ITS/030/21.
}
\thanks{Yiqun Lin, Jixiang Chen, Hualiang Wang, Jiewen Yang, Jiarong Guo, and Xiaomeng Li are with the Department of Electronic and Computer Engineering, the Hong Kong University of Science and Technology, Hong Kong SAR, China. Email: \{ylindw, jchenhu, hwangfd, jyangcu, jguoaz\}@connect.ust.hk, eexmli@ust.hk}
\thanks{Yi Zhang is with School of Cyber Science and Engineering, Sichuan University, Sichuan, China. Email: yzhang@scu.edu.cn}
\thanks{Xiaomeng Li is the corresponding author. Email: eexmli@ust.hk}
}

\maketitle

\newcommand{\ie}{\textit{i.e.}}
\newcommand{\eg}{\textit{i.e.}}
\newcommand{\vs}{\textit{v.s.}}
\newcommand{\etal}{\textit{et al.}}
\newcommand{\nickname}{DiCE}

\newcommand{\red}[1]{\textcolor{red}{#1}}
\newcommand{\green}[1]{\textcolor[rgb]{0.1,0.7,0.1}{#1}}

\newcommand{\xmli}[1]{\textcolor{red}{XM:#1}}
\newcommand{\cgs}[1]{#1}

\begin{abstract}

Cone-beam computed tomography (CBCT) is a critical 3D imaging technology in the medical field, while the high radiation exposure required for high-quality imaging raises significant concerns, particularly for vulnerable populations. Sparse-view reconstruction reduces radiation by using fewer X-ray projections while maintaining image quality, yet existing methods face challenges such as high computational demands and poor generalizability to different datasets. To overcome these limitations, we propose DeepSparse, the first foundation model for sparse-view CBCT reconstruction, featuring DiCE (Dual-Dimensional Cross-Scale Embedding), a novel network that integrates multi-view 2D features and multi-scale 3D features. Additionally, we introduce the HyViP (Hybrid View Sampling Pretraining) framework, which pretrains the model on large datasets with both sparse-view and dense-view projections, and a two-step finetuning strategy to adapt and refine the model for new datasets. Extensive experiments and ablation studies demonstrate that our proposed DeepSparse achieves superior reconstruction quality compared to state-of-the-art methods, paving the way for safer and more efficient CBCT imaging. The code will be publicly available at \href{https://github.com/xmed-lab/DeepSparse}{\tt\small https://github.com/xmed-lab/DeepSparse}.

\end{abstract}

\section{Introduction}

\noindent
Computed tomography (CT) is one of the most important imaging techniques in the medical field, enabling non-invasive visualization of internal anatomical structures in the human body. Based on the type of rays emitted, CT can be classified into fan/parallel-beam CT and cone-beam CT (CBCT). CBCT offers faster scanning speeds and improved resolution~\cite{scarfe2006clinical}. However, producing high-quality CT images requires hundreds of X-ray projections, resulting in significant radiation exposure to patients. This high radiation exposure can raise serious concerns in clinical practice~\cite{brenner2007computed, miglioretti2013use}, particularly for vulnerable populations such as pediatric patients and pregnant women~\cite{pearce2012radiation, lee2004diagnostic}. Therefore, reducing the number of X-ray projections while maintaining high-quality CT images is a promising solution to reduce the radiation dose. This approach is commonly referred to as sparse-view CT reconstruction.

\begin{figure}[t]
    \centering
    \includegraphics[width=1.0\linewidth]{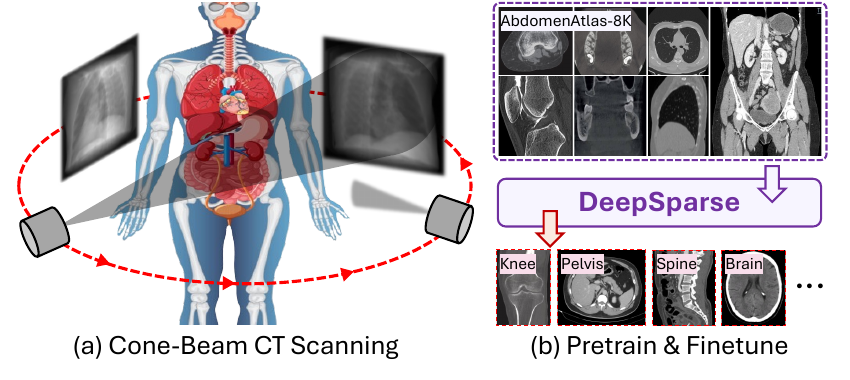}
    \vspace{-0.6cm}
    \caption{(a) During the CBCT scanning, the X-ray source will emit cone-shaped beams, and the measurement is a 2D projection at each view. (b) Our DeepSparse is pretrained on a large-scale CT dataset, covering various body organs with different projection parameters. DeepSparse can be further finetuned on target datasets to achieve the state-of-the-art reconstruction performance.}
    \label{fig:teaser}
\end{figure}

Previous research on sparse-view reconstruction has primarily focused on conventional fan/parallel-beam CT. Representative approaches include image-to-image translation methods~\cite{jin2017deep, han2016deep, guo2023self, zhang2018sparse, wang2018conditional, huang2018metal, ma2023freeseed}, which first reconstruct 2D low-quality CT slices from sparse 1D projections (sinograms) using filtered back projection (FBP) and then enhance the quality of these slices using 2D convolutional neural networks (CNNs). Additionally, some studies~\cite{wu2021drone, he2020radon, song2021solving, chung2023solving, lin2019dudonet, ma2023freeseed, wang2021dudotrans} explore reconstruction in the projection domain or both the projection and image domains. However, extending these methods to CBCT presents significant challenges: 1.) the increased dimensionality, because CBCT reconstruction involves reconstructing 3D volumes from multiple 2D X-ray projections (see Figure~\ref{fig:teaser}a), and 2.) the differences in the measurement processes~\cite{lin2023learning} between cone-beam and fan/parallel-beam.

In recent years, implicit neural representation~\cite{mildenhall2021nerf} has been introduced to CBCT reconstruction, where the 3D CT image is represented as a continuous attenuation coefficient field~\cite{lin2019dudonet, zha2022naf, lin2024learning, lin2024c2rv, fang2022snaf, shen2022nerp}. Leveraging this implicit representation, self-supervised methods like NAF~\cite{zha2022naf} propose to simulate the measurement process using digital reconstructed radiography (DRR) and minimize the error between measured and estimated projections. Although R$^\text{2}$-Gaussian~\cite{zha2024r} further accelerates the optimization by incorporating Gaussian Splatting, these methods still remain computationally expensive for per-sample optimization and struggle with extremely sparse projections.
Data-driven methods such as DIF-Net~\cite{lin2023learning} have been developed to train networks on a large dataset, learning a mapping from sparse projections to the attenuation field. Additionally, C$^\text{2}$RV~\cite{lin2024c2rv} further incorporates multi-scale 3D representations and cross-view attention to improve the reconstruction quality. However, these approaches face several limitations: 1.) the models are inefficient in scenarios with dense views, as the image encoder-decoder incurs increased computational costs with more input projections, 2.) a large amount of training data is required to achieve satisfactory reconstruction performance, and 3.) the trained models lack generalizability across different body parts, significantly limiting their practical applicability of these methods.

To address the limitations of previous data-driven reconstruction methods, we propose the first foundation model, namely DeepSparse (Figure~\ref{fig:teaser}b), for data-driven sparse-view CBCT reconstruction.
Firstly, the basic reconstruction network, DiCE (Dual-Dimensional Cross-Scale Embedding), is built upon C$^\text{2}$RV~\cite{lin2024c2rv} by removing the 2D decoder and introducing multi-scale projection encoding along with cross-scale 3D feature embedding. This design mitigates the computational overhead associated with an increased number of input views while maintaining comparable performance when trained from scratch.
Then, we observe that more accurate features would lead to superior reconstruction performance. Therefore, the development of DeepSparse is guided by two key objectives: 1.) pretraining the network on a large-scale dataset to improve the generalizability of the 2D encoder, and 2.) enhancing the 3D features through a denoising layer for feature refinement.
Specifically, we introduce the HyViP (Hybrid View Sampling Pretraining) framework, which pretrains the model using both sparse-view and dense-view projections to generate 2D and 3D features, respectively. Following this, we propose an effective two-step finetuning strategy: the first step adapts the pretrained model to a new target dataset, and the second step trains a denoising layer to refine the 3D features generated from sparse-view projections. Extensive experiments demonstrate that our proposed DeepSparse achieves significantly superior reconstruction performance compared to previous state-of-the-art methods across various datasets.

To summarize, the main contributions are as follows
\begin{itemize}
    \item DeepSparse, the first foundation model for sparse-view cone-beam CT reconstruction.
    \item DiCE, a novel CBCT reconstruction network to efficiently incorporate multi-scale projection encoding and cross-scale 3D feature embedding.
    \item HyViP, an innovative pretraining framework to improve the generalizability of the reconstruction model, and two-step finetuning to effectively adapt the pretrained model to various target datasets.
    \item Experiments and ablation studies are conducted to analyze the effectiveness of our proposed methods.
\end{itemize}

\section{Related Work}

\noindent
Sparse-view CBCT reconstruction presents unique challenges compared to traditional CT reconstruction due to its 3D nature and differing measurement geometries. This section reviews prior work in three areas: conventional fan/parallel-beam CT reconstruction, sparse-view cone-beam CT reconstruction, and the development of foundation models in medical imaging.

\subsection{Fan/Parallel-Beam CT Reconstruction}

\noindent
Conventional low-dose CT reconstruction methods are mainly proposed for parallel-beam or fan-beam CT, where the target is to restore a 2D slice from undersampled 1D sinograms. Previously, the reconstruction problem was formulated as an image-to-image translation~\cite{jin2017deep, han2016deep, zhang2018sparse, wang2018conditional, huang2018metal, ma2023freeseed} (also known as image-domain methods), where the low-quality CT slice is first reconstructed from undersampled projections by applying filtered back projection (FBP). Then a CNN-based network (\eg, U-Net~\cite{ronneberger2015u} and DenseNet~\cite{huang2017densely}) is employed to refine the low-quality slice. On the other hand, projection-domain methods are developed to recover missing sinograms~\cite{wu2021drone} or learn a mapping from 1D sinograms to the 2D slice~\cite{he2020radon, song2021solving, chung2023solving}. Furthermore, some works~\cite{lin2019dudonet, ma2023freeseed, wang2021dudotrans} have been proposed to leverage complementary information from both image and projection domains. 
Although these methods demonstrate considerable performance in conventional CT reconstruction, adapting them to cone-beam CT reconstruction poses significant challenges, such as much higher computational costs (due to increased dimensionality) and differences in measurement processes.

\subsection{Cone-Beam CT Reconstruction}

\noindent
Traditionally, the FDK algorithm~\cite{feldkamp1984practical} was developed as an extension of FBP to accommodate cone-beam geometries in 3D imaging. To handle sparse (50-100 views) and noisy data, ART-based methods~\cite{gordon1970algebraic, andersen1984simultaneous, pan2006variable} were introduced, employing an iterative approach to minimize the error between measured and estimated projections. However, these methods are sensitive to the initial state and often suffer from severe streaking artifacts when the number of views further decreases ($\leq$50 views).


\cgs{Recently, with the advancement of deep learning techniques in medical imaging, 3D CNNs-based methods~\cite{jiang2022mfct, shen2019patient, ying2019x2ct, kyung2023perspective} have been proposed for reconstruction in specific experimental settings, such as single-views~\cite{shen2019patient}, orthogonal-view~\cite{shen2019patient}, or patient-specific~\cite{shen2019patient} reconstruction. However, these methods are often limited by their specialized designs, such as reliance on fixed geometries~\cite{jiang2022mfct, ying2019x2ct, kyung2023perspective} or patient-specific training~\cite{shen2019patient}. Furthermore, they typically utilize heavy volumetric representations and 3D CNNs for feature encoding/decoding. Extending these architectures to general sparse-view scenarios (\eg, 6-10 views) would incur prohibitively high computational costs, limiting their practical applicability.}

Emerging works utilize implicit neural representations~\cite{mildenhall2021nerf, ruckert2022neat} to represent CBCT as a continuous attenuation~\cite{fang2022snaf, zha2022naf, shen2022nerp, lin2023learning, lin2024learning, lin2024c2rv}. Self-supervised methods, such as NAF~\cite{zha2022naf}, NeRP~\cite{shen2022nerp}, simulate the measurement process and minimize the error between real and synthesized projections (similar to ART-based methods~\cite{gordon1970algebraic, andersen1984simultaneous, pan2006variable}). R$^\text{2}$-Gaussian~\cite{zha2024r} further incorporates the Gaussian Splatting to accelerate the reconstruction speed. However, self-supervised approaches often require time-consuming per-sample optimization and perform poorly with extremely sparse views ($\leq$10 views) due to the lack of prior knowledge. 
In contrast, data-driven methods like DIF-Net~\cite{lin2023learning} and DIF-Gaussian~\cite{lin2024learning} aim to aim to learn a mapping from extremely sparse projections to the attenuation field from a large dataset. Additionally, C$^\text{2}$RV~\cite{lin2024c2rv} integrates multi-view 3D representation and cross-view attention to enhance the reconstruction quality. Nonetheless, the adaptation ability of these data-driven methods is limited, as the well-trained models require retraining to adapt to a new dataset, which significantly hinders their practical applicability.

\subsection{Foundation Models in Medical Imaging}

\noindent
Recently, foundation models have emerged as transformative tools in medical imaging~\cite{alkaeed2025open}, offering unprecedented capabilities across various domains. Specifically, foundation models such as Triad~\cite{wang2025triad} and Merlin~\cite{blankemeier2024merlin} have been developed for 3D MRI and CT imaging, respectively, leveraging vision-language architectures to enhance the understanding and interpretation of 3D medical data. For chest X-ray analysis, EVA-X~\cite{yao2024eva} and CheXagent~\cite{chen2024chexagent} employ self-supervised learning and advanced interpretability techniques to improve diagnostic accuracy and reliability.  
Foundation models have also shown promise in downstream applications such as medical report generation. For example, Li \etal~\cite{li2025towards} develop a multimodal framework that integrates 3D brain CT data with large language models, enabling automated, high-quality radiology report generation. Similarly, the granular alignment algorithm presented in~\cite{huang2024enhancing} aligns radiographic image representations with textual descriptions, enhancing the precision and coherence of radiology reports.  
In medical reconstruction, Terris \etal~\cite{terris2025reconstruct} propose a non-iterative, lightweight architecture that incorporates knowledge of the forward operator, demonstrating robust performance in denoising low-quality CT and undersampled MRI. 
\cgs{Similarly, Liu et al.~\cite{liu2024imaging} propose TAMP, an imaging foundation model pre-trained on large-scale physics-driven simulations to universally enhance CT images from non-ideal measurements.}
Additionally, the foundation model proposed in~\cite{liu2025foundation} leverages Gaussian Splatting and distills 3D cues from multiple vision foundation models to enable 4D dynamic scene reconstruction of deformable tissues, accurately capturing temporal changes in complex anatomical structures.

Despite these advancements, there is currently no foundation model designed specifically for sparse-view CBCT reconstruction. Developing such a model is crucial to improving both the generalization and adaptation capabilities for this challenging task, which remains a significant gap in the field of medical imaging.
\section{Method}

\noindent
In this section, we formally introduce DeepSparse, a foundation model for sparse-view CBCT reconstruction. We first revisit C$^\text{2}$RV~\cite{lin2024c2rv} and present the core reconstruction network DiCE, followed by a detailed explanation of HyViP, our pretraining approach with hybrid view sampling, and the two-step fine-tuning process.

\subsection{\cgs{Revisit C$^\text{2}$RV~\cite{lin2024c2rv}}}

\noindent
\cgs{Following previous works~\cite{lin2023learning, lin2024c2rv}, we define CT as a continuous implicit function $g: \mathbb{R}^3 \rightarrow \mathbb{R}$. This function maps a point $p \in \mathbb{R}^3$ in 3D space to its corresponding attenuation coefficient value $v \in \mathbb{R}$, \ie, $v = g(p)$. Given $N$-view projections $\mathcal{I} = \{I^1, \dots, I^N\} \subset \mathbb{R}^{W \times H}$ ($W$ and $H$ represent the width and height of the projections, respectively), the reconstruction problem is formulated as learning a conditional continuous implicit function $g(\cdot)$ such that $v = g(\mathcal{I}, p)$.}

\cgs{C$^\text{2}$RV~\cite{lin2024c2rv} utilizes 2D encoders to extract multi-view 2D features $\mathcal{F} = \{\mathcal{F}^1, \dots, \mathcal{F}^N\} \subset \mathbb{R}^{C\times(W\times H)}$ and back-project multi-view 2D features at different intermediate scales into the 3D voxelized space using imaging parameters to obtain multi-scale 3D features $\{F^{3D}_i, \dots, F^{3D}_S\}$. Given a point $p$, its multi-view pixel-aligned features and multi-scale voxel-aligned features are interpolated from 2D and 3D features, respectively. Finally, a scale-view cross-attention module is proposed to adaptively aggregate these interpolated features to estimate the attenuation value.
While C$^2$RV performs effectively on specific datasets, it faces significant limitations: it relies on large-scale training data, lacks generalizability across different anatomies, and cannot process dense-view inputs due to the high computational complexity of its network architecture. In this study, we propose DiCE, a more effective and efficient reconstruction framework built upon C$^2$RV. We then introduce HyViP pretraining with hybrid view sampling (combining sparse and dense inputs) with a 2-step finetuning process to learn universal anatomical and geometric representations from large-scale datasets.
}

\begin{figure*}[t]
    \centering
    \includegraphics[width=1.0\linewidth]{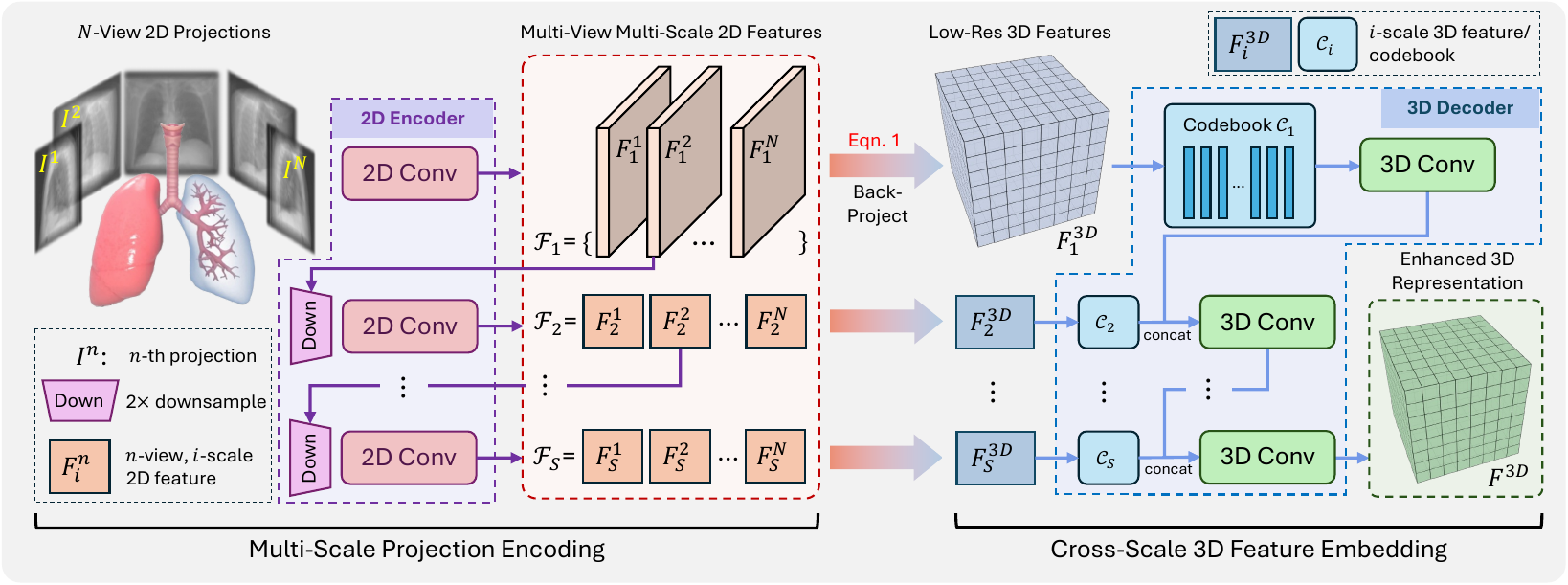}
    \caption{Overview of the reconstruction framework DiCE. The 2D encoder extracts multi-scale semantic features from sparse-view 2D projections. At each scale, these multi-view features are back-projected into a low-resolution volumetric space to generate 3D features. The 3D decoder then aggregates the multi-scale 3D features to produce an enhanced 3D representation.}
    \label{fig:framework}
\end{figure*}

\subsection{Reconstruction Framework --- \nickname{}}

\noindent
Firstly, we propose a more effective reconstruction framework \nickname{} (\underline{\textbf{D}}ual-D\underline{\textbf{i}}mensional \underline{\textbf{C}}ross-Scale \underline{\textbf{E}}mbedding) built upon C$^\text{2}$RV~\cite{lin2024c2rv}, which comprises three key components: a 2D encoder to extract multi-scale semantic features from input projections, a 3D decoder to generate a low-resolution 3D volumetric representation using back-projected features, and a point decoder that predicts the attenuation coefficient values for sampled points. An overview of DiCE is shown in Figure~\ref{fig:framework}.

\vspace{3pt}
\noindent
\textbf{Multi-Scale Projection Encoding.} A 2D encoder $\mathcal{E}(\cdot)$ with several convolutional and downsampling layers is used to extract semantic features from input sparse-view projections, producing multi-scale 2D features noted as $\mathcal{E}(I^n) = \{F_1^n, \dots, F_S^n\}$ for different views $n \in \{1, \dots, N\}$, where $S$ is the number of scales. Then, we denote those multi-scale multi-view features as $\{\mathcal{F}_1, \dots, \mathcal{F}_S\}=\mathcal{E}(\mathcal{I})$, where $\mathcal{F}_i = \{F^1_i, \dots, F^N_i\}$ for $i \in \{1, \dots, S\}$.

\vspace{3pt}
\noindent
\textbf{Low-Res 3D Features.} For each scale $i$, we follow C$^\text{2}$RV~\cite{lin2024c2rv} to back-project multi-view features $\mathcal{F}_i$ into the volumetric space, generating a 3D volumetric feature $F^\text{3D}_i \in\mathbb{R}^{c\times(r\times r\times r)}$. Specifically, the volumetric space $\mathcal{S} \subset \mathbb{R}^{3 \times (r \times r \times r)}$ is constructed by voxelizing the 3D space with a specific resolution $r$. \cgs{For each voxel $q$ in $\mathcal{S}$, we interpolate a set of pixel-aligned features $\big\{\hat{F}_i^n(q), \dots, \hat{F}_i^N(q)\big\}$ from multi-view 2D features at scale $i$, and then apply a max-pooling layer to aggregate them as the representative feature $F_i^{\text{3D}}(q)$ of voxel $q$. Specifically,}
\begin{equation}
\begin{aligned}
    F^\text{3D}_i(q) &= \text{Max-Pooling}\Big(\hat{F}^1_i(q), \dots, \hat{F}^N_i(q)\Big), \\
    \text{where}\ \hat{F}^n_i(q) &= \text{Interp}\Big(F^n_i, \pi^n(q)\Big)\text{, for } n \in \{1, \dots, N\},
\end{aligned}
\end{equation}
$\text{Interp}$:~$(\mathbb{R}^{C \times (D_1 \times \cdots \times D_k)}, \mathbb{R}^k) \rightarrow \mathbb{R}^C$ is $k$-linear interpolation, and $\pi^n: \mathbb{R}^3 \rightarrow \mathbb{R}^2$ is projection function of $n$-th view. Here, we denote the back-projection process as $\{F^\text{3D}_1, \dots, F^\text{3D}_S\} = \mathcal{B}\big(\{\mathcal{F}_1, \dots, \mathcal{F}_S\}\big)$.

\begin{figure}[t]
    \centering
    \includegraphics[width=0.8\linewidth]{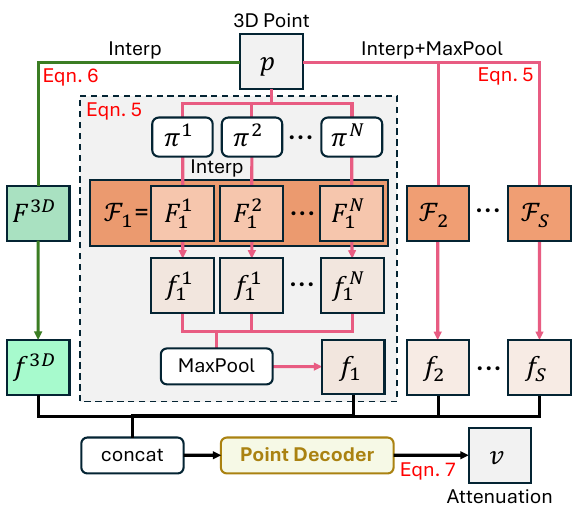}
    \vspace{-0.3cm}
    \caption{Attenuation prediction. For a sampled 3D point, we obtain its multi-scale pixel-aligned features from multi-view multi-scale 2D features by projection, interpolation, and applying max-pooling. Similarly, we obtain the voxel-aligned features through interpolation. Then, these features are concatenated and passed into the point decoder, predicting the corresponding attenuation coefficient for the point.}
    \label{fig:pdecoder}
\end{figure}

\vspace{3pt}
\noindent
\textbf{Cross-Scale 3D Feature Embedding.} Rather than directly utilizing the multi-scale 3D features $\{F^\text{3D}_1, \dots, F^\text{3D}_S\}$, we introduce a 3D decoder $\mathcal{D}(\cdot)$ to aggregate them, producing the enhanced 3D representation: \label{sec:codebook}
\begin{equation}
    F^\text{3D} = \mathcal{D}(\{F^\text{3D}_1, \dots, F^\text{3D}_S\}).
\end{equation}
To be more specific, 
\begin{equation}
    \tilde{F}_i^\text{3D} = \mathcal{C}_i(F^\text{3D}_i), \text{ and}
    \label{eq:codebook}
\end{equation}
\begin{equation}
    \hat{F}^\text{3D}_i = \left\{
    \begin{aligned}
        & \text{conv3D}\big(\tilde{F}_i^\text{3D}\big), i=1 \\[0.3em]
        & \text{conv3D}\Big(\big[\tilde{F}_i^\text{3D}, \hat{F}^\text{3D}_i\big]\Big), i \geq 2
    \end{aligned}
    \right.
\end{equation}
and $F^\text{3D} = \hat{F}^\text{3D}_S$, where $[\cdot, \cdot]$ indicates feature concatenation and $\mathcal{C}_i$ is the codebook at $i$-th scale for vector quantization. Here, $\mathcal{C}_i = \sigma_i^\text{post}\circ q_i \circ \sigma_i^\text{pre}$, where $\sigma_i$ are pre-/post-quantization linear layers and $q_i$ indicates feature quantization. Particularly, codebooks are introduced to capture the feature distribution of 3D features in the latent space, which will be explained further in the section detailing the finetuning steps.

Compared to C$^\text{2}$RV~\cite{lin2024c2rv}, we only use downsampling layers in the 2D encoder, with the decoding applied only to the 3D features. This design is more memory-efficient when handling a large number of views, as the resolution of the 2D features remains low. Additionally, the decoding process is agnostic to the number of views, as it operates on 3D features back-projected from the multi-view features.

\vspace{3pt}
\noindent
\textbf{Point Decoder.} Given a point $p\in \mathbb{R}^3$ defined over the 3D space, we firstly query its pixel-aligned features from multi-scale 2D features $\{\mathcal{F}_1, \dots, \mathcal{F}_S\}$, where $\mathcal{F}_i = \{F_i^1, \dots, F_i^N\}$. As shown in Figure~\ref{fig:pdecoder}, for each scale:
\begin{equation}
\begin{aligned}
    f_i &= \text{MaxPooling}\Big(\big\{f_i^1, \dots, f_i^N\big\}\Big), \\
    \text{where } f_i^n &= \text{Interp}\Big(F^n_i, \pi^n(p)\Big), \text{ for $n\in \{1, \dots, N\}$}.
\end{aligned}
\label{eq:interp2d}
\end{equation}
Then we query its voxel-aligned features from low-res 3D volumetric representations $F^\text{3D}$:
\begin{equation}
    f^\text{3D} = \text{Interp}(F^\text{3D}, p).
\end{equation}
Finally, we concatenate all the queried features mentioned above and utilize MLPs to predict the attenuation coefficient value for the point:
\begin{equation}
    v = \text{MLPs}\Big(\big[f_1, \dots, f_S, f^\text{3D}\big]\Big).
    \label{eq:mlp}
\end{equation}

\subsection{Model Pretraining -- HyViP}

\noindent
In the proposed reconstruction framework \nickname{}, there are two types of features encoded from the input projections: multi-view 2D features and a 3D feature. These features are used to predict point-wise attenuation coefficients, where more accurate features lead to improved reconstruction quality. 
To enhance 2D features, the model can be pretrained on a large-scale dataset to improve the generalization capability of the 2D encoder. Additionally, we formulate the generation of more accurate 3D features from sparse-view projections as a feature-denoising problem. To be more specific, vector quantization ($\mathcal{C}$ in Eqn.~\ref{eq:codebook}) is incorporated into the 3D decoder, allowing the learning of high-quality codebook priors from dense-view projections during pretraining. Then, a feature denoising layer is introduced to refine the sparse-view features, aligning them with dense-view features during finetuning.

\begin{figure}[t]
    \centering
    \includegraphics[width=0.8\linewidth]{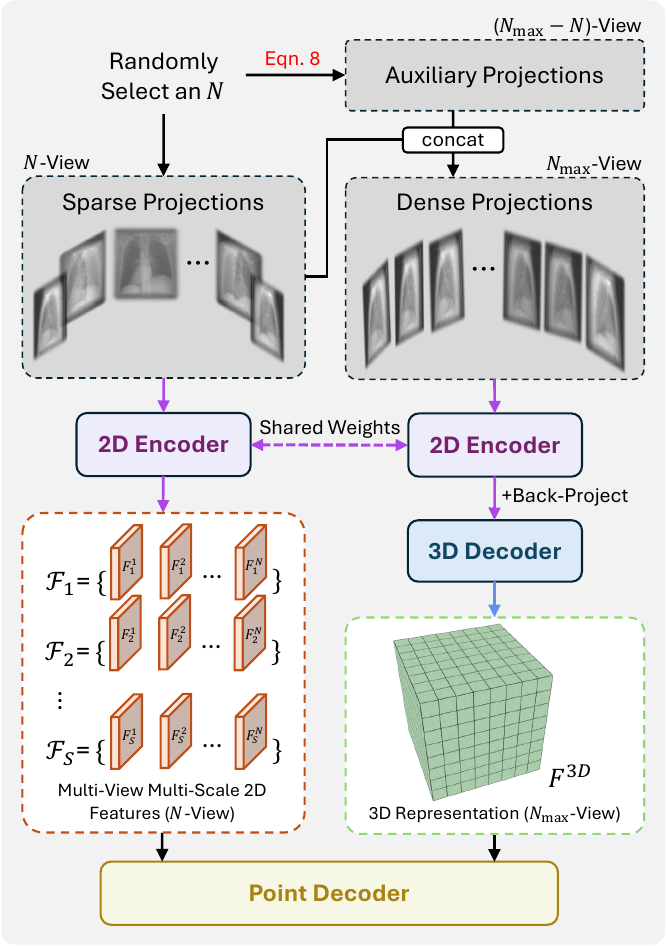}
    \caption{Overview of HyViP pretraining framework. In each training iteration, we randomly select an $N$ and sample $N$-view sparse projections and $N_\text{max}$-view dense projections, which are then used to generate multi-view multi-scale 2D features and 3D representation, respectively.}
    \label{fig:pretrain}
\end{figure}

In this subsection, we formally introduce HyViP (\underline{\textbf{Hy}}brid \underline{\textbf{Vi}}ew Sampling \underline{\textbf{P}}retraining) to pretrain the model using hybrid view sampling methods. The overview of HyViP pretraining framework is shown in Figure~\ref{fig:pretrain}. Specifically, we define the minimum and maximum number of views $N_\text{min}$ and $N_\text{max}$. 
In each training iteration, we randomly select an integer $N \in [N_\text{min}, N_\text{max}]$. We first uniformly sample $N$ and $N_\text{max}$ viewing angles as $\Lambda_\text{sparse}$ and $\Lambda_\text{dense}$, respectively. Then, we find a set of auxiliary viewing angles $\Lambda_\text{aux}$ to 
\begin{equation}
\begin{aligned}
    \text{minimize:\ }&d(\Lambda, \Lambda_\text{dense}) = \min_{\phi: \Lambda \rightarrow \Lambda_\text{dense}} \sum_{\alpha \in \Lambda} \big\|\alpha - \phi(\alpha)\big\|_1 \\
    \text{s.t. } & \left\{
    \begin{aligned}
        &\Lambda_\text{aux} \subset \Lambda_\text{dense} \\
        &|\Lambda_\text{aux}| = N_\text{max} - N \\
        &\Lambda_\text{sparse} \cap \Lambda_\text{aux}  = \emptyset
    \end{aligned}
    \right.
\end{aligned}
\end{equation}
where $\Lambda = \Lambda_\text{sparse} \cup \Lambda_\text{aux}$, and $\phi$ indicates the bijection between $\Lambda$ and $\Lambda_\text{dense}$. Then, we sample $N_\text{max}$ projections $\mathcal{I}$ corresponding to the viewing angles in $\Lambda$. Particularly, we denote $\mathcal{I}_{1:N}$ as the first $N$ projections of $\mathcal{I}$ corresponding to the viewing angles $\Lambda_\text{sparse}$. During training, we use the first $N$ projections $\mathcal{I}_{1:N}$ to generate multi-view multi-scale 2D features $\{\mathcal{F}_1, \dots, \mathcal{F}_S\} = \mathcal{E}(\mathcal{I}_{1:N})$, and all $N_\text{max}$ projections $\mathcal{I}$ to generate the 3D representation $F^\text{3D}=\mathcal{D}\circ\mathcal{B}\circ\mathcal{E}(\mathcal{I})$. Then, the prediction (Eqn.~\ref{eq:interp2d}-\ref{eq:mlp}) for the attenuation coefficient of a sampled point is calculated based on the above $\{\mathcal{F}_1, \dots, \mathcal{F}_S\}$ ($N$-view) and $F^\text{3D}$ ($N_\text{max}$-view).

The training loss includes the task loss and quantization loss, \ie, $\mathcal{L} = \mathcal{L}_\text{task} + \lambda_1\cdot\mathcal{L}_\text{vq}$, where $\lambda_1$ is the scaling factor to control the trade-off. Specifically, we follow \cite{lin2024c2rv, lin2023learning} to use mean squared error (MSE) as the task loss to measure reconstruction error between predicted attenuation coefficients ($v_j$) and ground-truth values ($\hat{v}_j$):
\begin{equation}
    \mathcal{L}_\text{task} = \sum_{j=1}^{N_p}(v_j - \hat{v}_j)^2,
\end{equation}
where $N_p$ is the number of sampled points. The quantization loss is used to penalize the difference between continuous features and their quantized representations:
\begin{equation}
    \mathcal{L}_\text{vq} = \sum_{i=1}^S \Big\|\dot{F}^\text{3D}_i - \text{sg}\big[q_i(\dot{F}^\text{3D}_i)\big]\Big\|_2^2,
\end{equation}
where $\dot{F}^\text{3D}_i = \sigma_i^\text{pre}(F^\text{3D}_i)$ and $\text{sg}[\cdot]$ indicates stopping gradient propagation. We follow \cite{razavi2019generating} to update codebook features of $\mathcal{C}_i$ via EMA (Exponential Moving Average). 

Our proposed HyViP pretraining the model on large-scale data with hybrid view sampling methods improves the model's generalization capability and enhances its robustness to variations in the number of projection views. Consequently, the model can be pretrained once and subsequently finetuned to accommodate diverse datasets and experimental settings.

\begin{figure*}[t]
    \centering
    \includegraphics[width=0.8\linewidth]{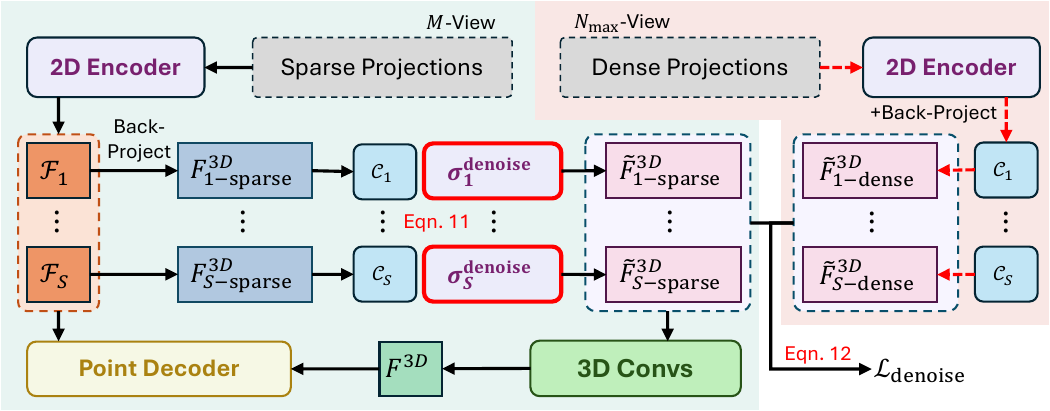}
    \caption{Overview of finetuning step-2. For sparse inputs, additional denoise layers ($\sigma^\text{denoise}_i$) are introduced to refine the quantized 3D features. For dense inputs, we stop the gradient propagation and use the quantized 3D features as a supervision to compute the denoise loss $\mathcal{L}_\text{denoise}$. Finally, only features (\ie., $\mathcal{F}_i$ and $F^\text{3D}$) generated from sparse projections are used to predict the attenuation coefficients.}
    \label{fig:finetune}
\end{figure*}

\subsection{Model Finetuning}

\noindent
During the model pretraining, 2D features are generated from random $N$ views, while 3D features are generated from a fixed $N_\text{max}$ views, where $N \leq N_\text{max}$. Consequently, the pretrained model cannot be directly applied to general $N$-view reconstruction, as the quality of 3D features will deteriorate greatly when generated from $N$ views. Therefore, to adapt the pretrained model to a target dataset and a specific number of views (\eg, $M$), two finetuning steps are required: 1.) adapt the model to the new dataset, and 2) adjust the number of views used for generating 3D features to match the specific $M$. 

To concisely represent the view sampling in the finetuning steps, we use $\mathcal{N}_\text{rand}$ to indicate the number can be randomly selected between $N_\text{min}$ and $N_\text{max}$. Additionally, $(N_\text{2D}, N_\text{3D})$ represents that we sample $N_\text{2D}$ views to generate 2D features and $N_\text{3D}$ views to generate 3D features. For example, the view sampling of the pretraining can be expressed as $(\mathcal{N}_\text{rand}, N_\text{max})$.
Then, denote the parameters of the pretrained model as $\theta_\text{pt}$, and two steps are introduced in detail as follows.

\vspace{3pt}
\noindent
\textbf{Step-1: Dataset Adaptation.} 
The goal of this step is to adapt the pretrained model to the target dataset using a specific number of views (\ie, $M$) to generate 2D features. Formally, we finetune all model parameters $\theta_\text{pt} \rightarrow \theta_\text{ft}^1$ on the target dataset with view sampling $(M, N_\text{max})$. Here, the 2D encoder is not frozen, as the value range of projections may differ across datasets and experimental settings. Then, the finetuning loss is the same as the one introduced in the pretraining.

\vspace{3pt}
\noindent
\textbf{Step-2: View Adjustment.} 
Next, we finetune the model $\theta_\text{ft}^1$ to $\theta_\text{ft}^2$ by adjusting the view sampling from $(M, N_\text{max})$ to $(M, M)$. As mentioned in Section~\ref{sec:codebook}, codebooks are designed to capture the feature distribution of 3D features generated from $N_\text{max}$ views. Therefore, these codebook embeddings cannot be directly utilized when reducing $N_\text{max}$ to $M$, as the quality of the 3D features degrades with fewer input views. To address this, we formulate the finetuning as a feature-denoising problem, where the low-quality 3D features generated from $M$ (sparse) views are refined by a denoise layer to align with the high-quality 3D features generated from $N_\text{max}$ (dense) views. 

Specifically, as shown in Figure~\ref{fig:finetune}, we first sample $M$ sparse projections, denoted as $\mathcal{I}_\text{sparse}$, and then supplement $\mathcal{I}_\text{sparse}$ to $N_\text{max}$ projections (denoted as $\mathcal{I}_\text{dense}$) by sampling additional auxiliary projections, following a similar view sampling strategy as in pretraining. The network takes both $\mathcal{I}_\text{sparse}$ and $\mathcal{I}_\text{dense}$ as inputs to generate 3D features. At each scale $i$, we introduce a denoising layer, $\sigma_i^\text{denoise}$, to refine the quantized features of the sparse inputs. Consequently, the Eqn.~\ref{eq:codebook} is modified to the following form to account for sparse inputs.
\begin{equation}
    \tilde{F}^\text{3D}_{i\text{-sparse}} = \sigma_i^\text{denoise} \circ \mathcal{C}_i(F^\text{3D}_{i\text{-sparse}}),
\end{equation}
where $F^\text{3D}_{i\text{-sparse}}$ represents the back-projected volumetric features of sparse inputs. Similarly, $\tilde{F}^\text{3D}_{i\text{-dense}}$ denotes the quantized features of dense inputs. During the finetuning, an additional denoise loss term is included:
\begin{equation}
    \mathcal{L}_\text{denoise} = \sum_{i=1}^S\Big\|\tilde{F}^\text{3D}_{i\text{-sparse}} - sg\big[\tilde{F}^\text{3D}_{i\text{-dense}}\big]\Big\|_1.
\end{equation}
In this step, only features generated from sparse inputs are used to estimate attenuation coefficients, which subsequently contribute to the task loss. The quantization loss ($\mathcal{L}_\text{vq}$) is not required as the encoder and codebooks are frozen. Then, the overall loss function for this step is defined as:
\begin{equation}
    \mathcal{L} = \mathcal{L}_\text{task} + \lambda_2\cdot\mathcal{L}_\text{denoise},
\end{equation}
where $\lambda_2$ is the scaling factor for the trade-off. In practice, the denoise layer is implemented as a shallow 3D CNN.

\begin{table}[t]
    \caption{Training configurations of each stage. LR: learning rate. For encoder/codebook/decoder, \ding{51} indicates this part is trainable while \ding{55} means it is frozen in this stage.} \label{tab:config}
    \centering
    \setlength{\tabcolsep}{3pt}
    \resizebox{1.0\linewidth}{!}{
    \begin{tabular}{l|c|c|c}
    \toprule[1.2pt]
        \multirow{2}{*}{Stage} &  \multirow{2}{*}{Pretrain} & \multicolumn{2}{c}{Finetune} \\ \cline{3-4}
         & & Step-1 & Step-2 \\ \hline \hline
        View Sampling & $(N_\text{rand}, N_\text{max})$ & $(M, N_\text{max})$ & $(M, M)$ \\ \hline
        Dataset & AbdomenAtlas-8K~\cite{qu2023abdomenatlas} & Taget Set & Target Set \\ \hline
        \# Data & 8,407 & 600$\sim$800 & 600$\sim$800 \\ \hline
        Epochs & 1,000 & 200 & 200 \\ \hline
        Batch Size & 16 & 2 & 2 \\ \hline
        Optimizer & AdamW + LR=$10^{-4}$ & AdamW + LR=$10^{-4}$ & AdamW + LR=$10^{-4}$ \\ \hline
        Loss & $\mathcal{L}_\text{task} + \lambda_1\cdot\mathcal{L}_\text{vq}$ & $\mathcal{L}_\text{task} + \lambda_1\cdot\mathcal{L}_\text{vq}$ &  $\mathcal{L}_\text{task} + \lambda_2\cdot\mathcal{L}_\text{denoise}$ \\ \hline
        2D Encoder & \ding{51} & \ding{51} & \ding{55} \\ \hline
        Codebook & \ding{51} & \ding{51} & \ding{55} \\ \hline
        3D Decoder & \ding{51} & \ding{51} & \ding{51} \\ \hline
        Point Decoder & \ding{51} & \ding{51} & \ding{51} \\ \hline
        w/ Denoise Layer & \ding{55} & \ding{55} & \ding{51} \\
    \bottomrule[1.2pt]
    \end{tabular}
}
\end{table}

\begin{table*}[t]
{
\centering
\caption{Comparison of our method and previous self-supervised methods on two datasets (\ie, chest and knee) with various numbers of projection views. The resolution of the reconstructed CT is 256$^\text{3}$. The reconstruction results are evaluated with PSNR (dB) and SSIM ($\times$10$^\text{-2}$), where higher PSNR/SSIM indicate better performance. The best values are \textbf{bolded}.}
\label{tab:ssl}
\setlength{\tabcolsep}{10pt}
\resizebox{1.0\linewidth}{!}{
\begin{tabular}{l|ccc|ccc}
\toprule[1.2pt]
\multirow{2}{*}{Method} & \multicolumn{3}{c|}{LUNA16~\cite{setio2017validation} (Chest)} & \multicolumn{3}{c}{Lin \etal~\cite{lin2023learning} (Knee)} \\ \cline{2-7}
 & 6-View & 8-View & 10-View & 6-View & 8-View & 10-View \\ 
 \hline \hline
FDK~\cite{feldkamp1984practical} & 15.29$|$27.80 & 16.54$|$28.05 & 17.36$|$29.06 & 18.42$|$30.56 & 19.83$|$32.42 & 20.95$|$34.55 \\
SART~\cite{andersen1984simultaneous} & 21.57$|$61.26 & 22.80$|$66.24 & 23.76$|$69.48 & 24.30$|$64.88 & 25.23$|$68.28 & 25.97$|$70.79 \\
NAF~\cite{zha2022naf} & 18.76$|$39.02 & 20.51$|$46.09 & 22.17$|$52.57 & 20.11$|$47.35 & 22.42$|$55.19 & 24.26$|$61.72 \\
NeRP~\cite{shen2022nerp} & 23.55$|$60.59 & 25.83$|$67.81 & 26.12$|$69.42 & 24.24$|$56.78 & 25.55$|$61.56 & 26.33$|$67.70 \\
DeepSparse (\textit{Ours}) & \textbf{30.22}$|$\textbf{89.96} & \textbf{31.14}$|$\textbf{90.76} & \textbf{31.86}$|$\textbf{91.41} & \textbf{33.16}$|$\textbf{91.28} & \textbf{34.28}$|$\textbf{93.35} & \textbf{35.41}$|$\textbf{93.63} \\
\bottomrule[1.2pt]
\end{tabular}
}
}
\end{table*}

\subsection{Implementation}

\noindent
We implement the network and training processes using the PyTorch framework~\cite{imambi2021pytorch}. For the reconstruction network, we set the scale $S$ to 4 and use a volumetric resolution of $r=32$ in our experiments. During the pretraining and finetuning, $N_\text{min}$ and $N_\text{max}$ are chosen as 6 and 24, respectively. Empirically, we set the scaling factors $\lambda_1=0.1$ and $\lambda_2=1.0$ for the quantization loss and denoise loss, respectively, to achieve optimal performance. At all stages, the model parameters are optimized using the AdamW optimizer with a learning rate of $10^{-4}$. During pretraining, the model is trained for 1,000 epochs with a batch size of 16, utilizing 4 GeForce RTX 3090 GPUs. For each finetuning step, the model is trained on a single GeForce RTX 3090 GPU for 200 epochs with a batch size of 2. A detailed comparison of the training configurations for each stage is provided in Table~\ref{tab:config}. Additional implementation details will be made available when the code is released.

\section{Experiments}

\noindent
In this section, we first pretrain the reconstruction model (\ie, DiCE) using a large-scale public CT dataset AbdomenAtlas-8K~\cite{qu2023abdomenatlas} with HyViP pretraining. We then perform extensive experiments by finetuning the pretrained model on various target sets with different experimental settings. Furthermore, we conduct comprehensive ablation studies to validate the effectiveness of each proposed module and the robustness of the pretrained model in non-ideal scenarios.

\subsection{Experimental Setting}

\noindent
\textbf{Datsets.}
We pretrained the model on AbdomenAtlas-8K~\cite{qu2023abdomenatlas}, consisting of 5,195 CT covering \cgs{diverse anatomical structures  from head, chest, abdomen, pelvis, to knee.} AbdomenAtlas-8K was collected from 26 hospitals worldwide, ensuring diversity in imaging protocols and patient demographics. To further validate the robustness and effectiveness of the pretrained model, we finetune the model on various target datasets and experimental settings, 
including LUNA16~\cite{setio2017validation} (chest), Lin~\etal~\cite{lin2023learning} (knee), \cgs{ToothFairy~\cite{cipriano2022deep} (head), PANORAMA~\cite{alves2024panorama} (abdomen), and PENGWIN~\cite{liu2025automatic} (pelvis)}. Specifically, LUNA16~\cite{setio2017validation} contains 888 chest CT with train/val/test splits of 738/50/100; Lin~\etal~\cite{lin2023learning} contains 614 knee CBCT, split into 464/50/100; \cgs{ToothFairy~\cite{cipriano2022deep} contains 443 head CBCT, split into 343/25/75; PANORAMA~\cite{alves2024panorama} contains 2,044 abdominal CT, split into 1244/200/600; PENGWIN~\cite{liu2025automatic} contains 100 pelvic CT, split into 60/10/30.} 

\begin{table*}[t]
\centering
\caption{Comparison of data-driven methods on five CT datasets with various numbers of projection views. The resolution of the reconstructed CT is 256$^\text{3}$. The reconstruction results are evaluated with PSNR (dB) and SSIM ($\times$10$^\text{-2}$), where higher PSNR/SSIM indicate better performance. The best values are \textbf{bolded} and the second-best values are \underline{underlined}.} \label{tab:datadriven}
\setlength{\tabcolsep}{10pt}
\resizebox{0.85\linewidth}{!}{
\begin{tabular}{l|c|c|c|c|c}
\toprule[1.2pt]
\multicolumn{6}{c}{\textbf{\textit{6-View}}} \\
\midrule[1.2pt]
Method & \makecell{\cgs{ToothFairy~\cite{cipriano2022deep}} \\ \cgs{(Head)}} & \makecell{LUNA16~\cite{setio2017validation} \\ (Chest)} & \makecell{\cgs{PANORAMA~\cite{alves2024panorama}} \\ \cgs{(Abdomen)}} & \makecell{\cgs{PENGWIN~\cite{liu2025automatic}} \\ \cgs{(Pelvis)}} & \makecell{Lin \etal~\cite{lin2023learning} \\ (Knee)} \\
\midrule[1.2pt]
FBPConvNet~\cite{jin2017deep} & 26.90$|$74.09 & 24.38$|$65.97 & 23.71$|$72.21 & 22.60$|$54.49 & 25.10$|$72.07 \\
FreeSeed~\cite{ma2023freeseed} & 27.72$|$84.60 & 25.59$|$66.03 & 24.14$|$76.55 & 23.90$|$73.71 & 26.74$|$73.42 \\
BBDM~\cite{li2023bbdm} & 27.52$|$83.01 & 24.78$|$65.80 & 24.23$|$74.03 & 23.70$|$72.71 & 26.58$|$74.42 \\
PixelNeRF~\cite{yu2021pixelnerf} & 24.55$|$71.58 & 24.66$|$66.49 & 23.33$|$62.58 & 23.18$|$66.55 & 26.10$|$79.96 \\
DIF-Net~\cite{lin2023learning} & 25.56$|$74.08 & 25.55$|$73.19 & 23.64$|$64.03 & 23.21$|$64.81 & 27.12$|$80.74 \\
C$^\text{2}$RV~\cite{lin2024c2rv} & \underline{28.22}$|$\underline{86.93} & \underline{29.23}$|$\underline{87.47} & \underline{25.96}$|$\underline{81.16} & \underline{25.35}$|$\underline{78.96} & \underline{29.73}$|$\underline{88.87} \\
DeepSparse (\textit{Ours}) & \textbf{30.19}$|$\textbf{90.66} & \textbf{30.22}$|$\textbf{89.96} & \textbf{27.53}$|$\textbf{85.34} & \textbf{27.53}$|$\textbf{86.41} & \textbf{33.16}$|$\textbf{91.28} \\
\midrule[1.2pt]
\multicolumn{6}{c}{\textbf{\textit{8-View}}} \\
\midrule[1.2pt]
FBPConvNet~\cite{jin2017deep} & 27.85$|$81.14 & 24.87$|$67.21 & 24.72$|$76.13 & 24.15$|$67.56 & 25.93$|$72.86 \\
FreeSeed~\cite{ma2023freeseed} & 28.04$|$83.81 & 26.86$|$67.44 & 25.49$|$75.31 & 24.27$|$71.03 & 27.88$|$75.82 \\
BBDM~\cite{li2023bbdm} & 27.75$|$83.99 & 25.81$|$67.06 & 25.77$|$76.66 & 24.31$|$74.31 & 28.01$|$75.71 \\
PixelNeRF~\cite{yu2021pixelnerf} & 26.13$|$73.01 & 25.04$|$68.24 & 23.68$|$64.03 & 23.65$|$66.94 & 26.84$|$81.33 \\
DIF-Net~\cite{lin2023learning} & 26.04$|$75.33 & 26.09$|$76.96 & 24.21$|$75.06 & 23.91$|$70.12 & 28.31$|$82.03 \\
C$^\text{2}$RV~\cite{lin2024c2rv} & \underline{28.92}$|$\underline{87.86} & \underline{29.95}$|$\underline{88.46} & \underline{26.56}$|$\underline{82.49} & \underline{25.92}$|$\underline{80.60} & \underline{30.68}$|$\underline{89.96} \\
DeepSparse (\textit{Ours}) & \textbf{31.21}$|$\textbf{91.86} & \textbf{31.14}$|$\textbf{90.76} & \textbf{28.31}$|$\textbf{86.32} & \textbf{28.52}$|$\textbf{88.66} & \textbf{34.28}$|$\textbf{93.35} \\
\midrule[1.2pt]
\multicolumn{6}{c}{\textbf{\textit{10-View}}} \\ 
\midrule[1.2pt]
FBPConvNet~\cite{jin2017deep} & 28.01$|$82.21 & 25.90$|$68.98 & 25.82$|$80.32 & 25.02$|$75.32 & 26.74$|$75.51 \\
FreeSeed~\cite{ma2023freeseed} & 28.80$|$85.24 & 27.23$|$68.62 & 26.67$|$78.04 & 25.39$|$75.37 & 28.77$|$77.87 \\
BBDM~\cite{li2023bbdm} & 28.70$|$85.34 & 26.35$|$68.71 & 26.82$|$78.97 & 25.66$|$75.69 & 28.90$|$77.26 \\
PixelNeRF~\cite{yu2021pixelnerf} & 26.63$|$75.33 & 25.39$|$70.62 & 24.16$|$65.79 & 23.89$|$67.65 & 27.36$|$82.49 \\
DIF-Net~\cite{lin2023learning} & 27.63$|$83.14 & 26.69$|$78.56 & 26.73$|$77.84 & 24.32$|$67.49 & 29.33$|$84.98 \\
C$^\text{2}$RV~\cite{lin2024c2rv} & \underline{29.67}$|$\underline{88.73} & \underline{30.70}$|$\underline{89.16} & \underline{27.23}$|$\underline{82.97} & \underline{26.54}$|$\underline{81.54} & \underline{31.55}$|$\underline{90.83} \\
DeepSparse (\textit{Ours}) & \textbf{31.79}$|$\textbf{92.50} & \textbf{31.86}$|$\textbf{91.41} & \textbf{29.42}$|$\textbf{88.36} & \textbf{29.03}$|$\textbf{90.27} & \textbf{35.41}$|$\textbf{93.63} \\
\bottomrule[1.2pt]
\end{tabular}
}
\end{table*}
\begin{figure*}[t]
    \centering
    \includegraphics[width=0.85\linewidth]{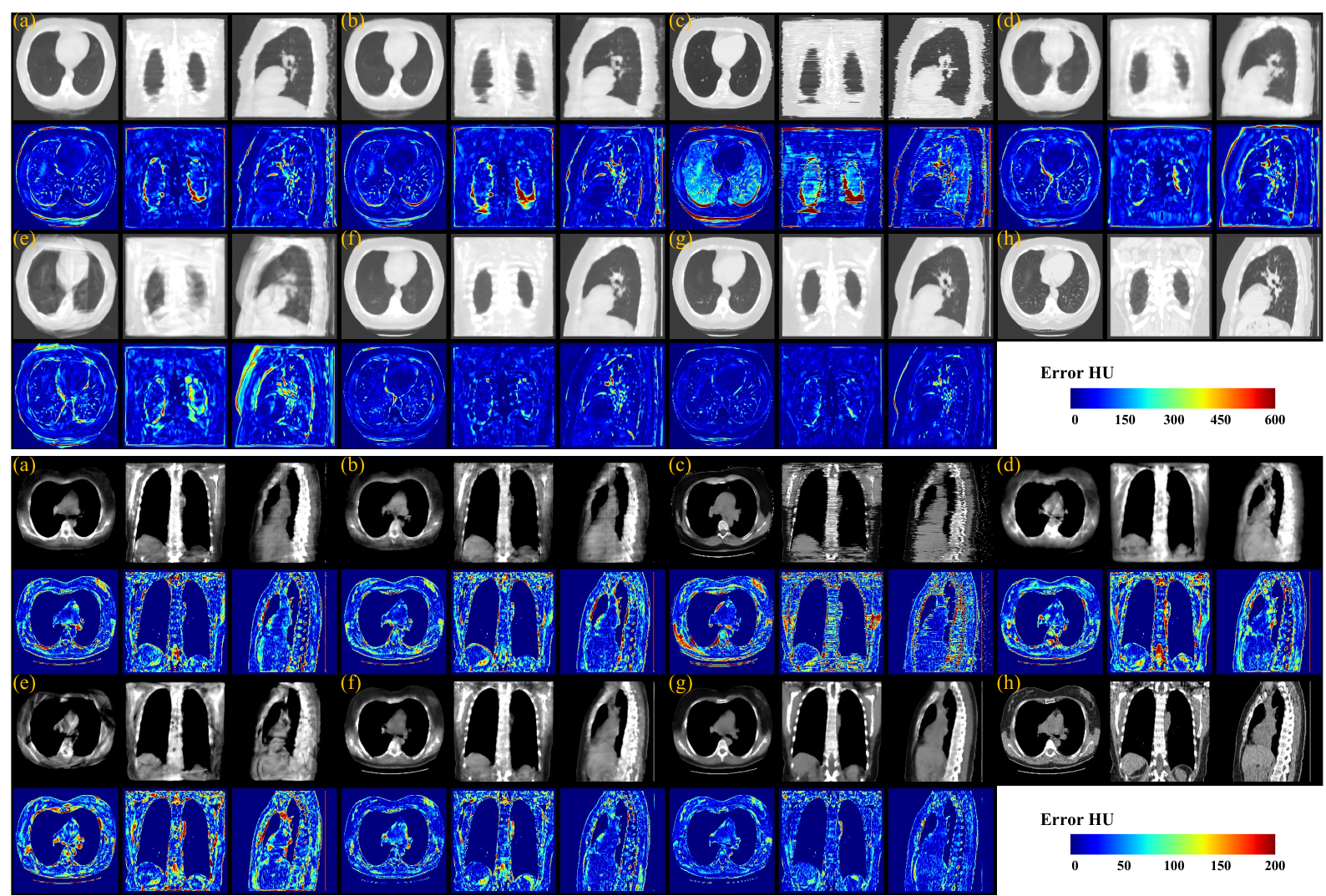}
    \vspace{-0.2cm}
    \caption{Visualization of reconstructed CT \cgs{and error maps}. Experiments are conducted on the LUNA16~\cite{setio2017validation} dataset with 6 projection views. \cgs{(a) FBPConvNet~\cite{jin2017deep}, (b) Freeseed~\cite{ma2023freeseed}, (c) BBDM~\cite{li2023bbdm}, (d) PixelNeRF~\cite{yu2021pixelnerf}, (e) DIF-Net~\cite{lin2023learning}, (f) C$^\text{2}$RV~\cite{lin2024c2rv}, (g) Ours, (h) Ground Truth. The top display is configured with a window level of -600 and width of 1500, while the bottom display is set to a level of 40 and width of 350.}}
    \label{fig:vislung}
\end{figure*}

\begin{figure*}[t]
    \centering
    \includegraphics[width=0.85\linewidth]{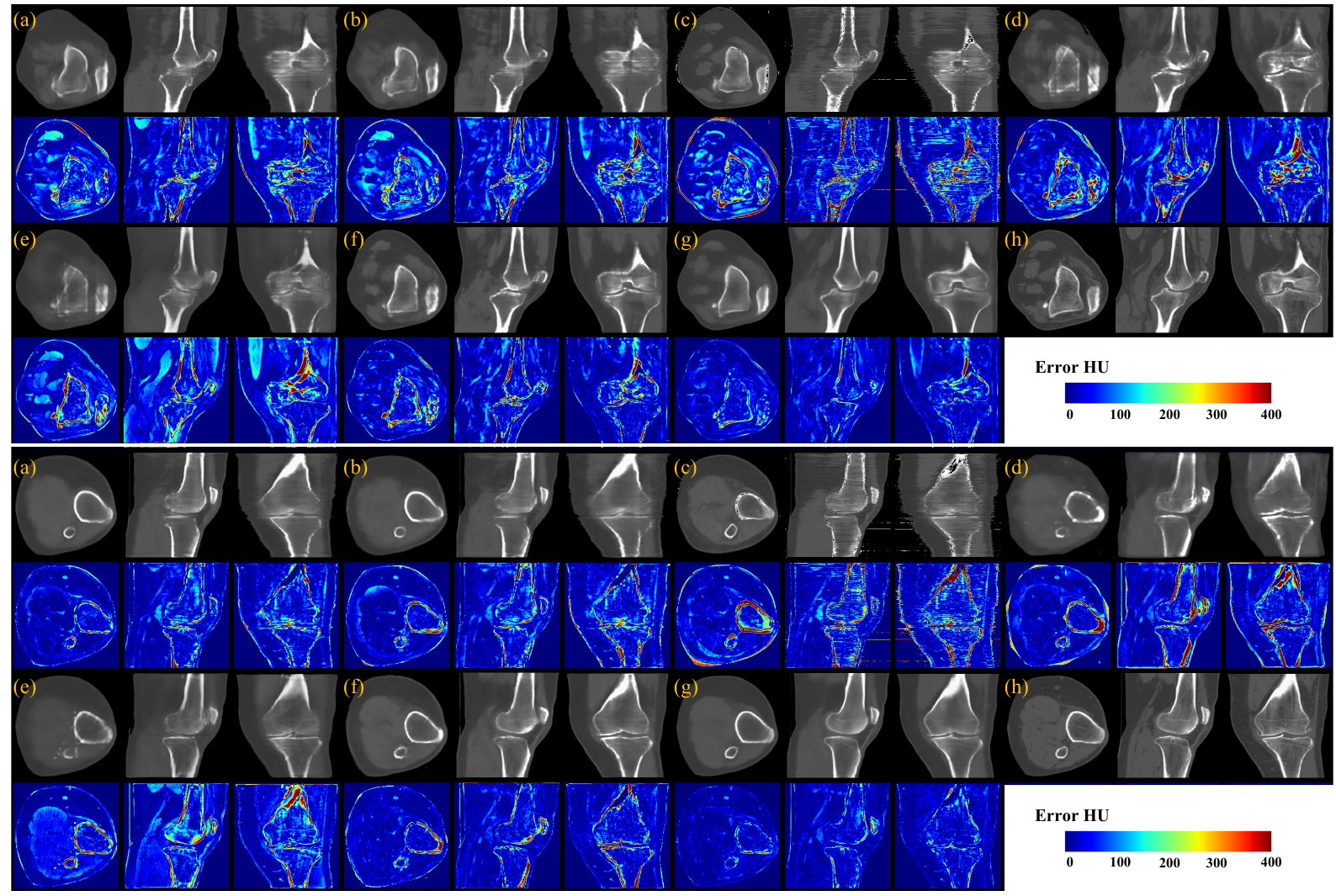}
    \vspace{-0.2cm}
    \caption{Visualization of reconstructed CT \cgs{and error maps}. Experiments are conducted on the knee dataset (Lin~\etal~\cite{lin2023learning}) with 6 projection views. \cgs{(a) FBPConvNet~\cite{jin2017deep}, (b) Freeseed~\cite{ma2023freeseed}, (c) BBDM~\cite{li2023bbdm}, (d) PixelNeRF~\cite{yu2021pixelnerf}, (e) DIF-Net~\cite{lin2023learning}, (f) C$^\text{2}$RV~\cite{lin2024c2rv}, (g) Ours, (h) Ground Truth. The display is configured with a window level of 300 and width of 1500.}}
    \label{fig:visknee}
\end{figure*}

\vspace{3pt}
\noindent
\textbf{Data Preprocessing.} 
The maximum length of AbdomenAtlas-8K in the axial direction is 1,983 mm, which is significantly larger than other datasets. Therefore, we crop the CT data into sub-volumes with a fixed size of 384 mm in the axial direction instead of applying center-cropping or resizing, resulting in a total of 8,407 CT volumes. Following \cite{lin2024c2rv}, we preprocess the 3D CT data and generate 2D X-ray projections from the CT volumes using digital reconstruction radiography (DRRs). 
\cgs{For AbdomenAtlas-8K, we utilize a standard geometric configuration with a Distance Source-to-Object of 800 mm and a Distance Source-to-Detector of 1200 mm. The volume size is resampled to $256^3$ with an isotropic voxel spacing of 1.5 mm. To model realistic noise, we apply Poisson distributed quantum noise (incident photon flux $I_0 = 1 \times 10^5$) followed by Gaussian noise ($\sigma=10$).}
Rather than generating sparse-view projections during training, we pre-generate 200 candidate projections with viewing angles uniformly sampled in $[0^\circ, 180^\circ]$ (half rotation) and load the corresponding projections based on the selected sparse viewing angles during training.
\cgs{Geometric parameters for target datasets are adjusted to align with their respective clinical protocols. Detailed configuration for all datasets will be released with our code.}

\vspace{3pt}
\noindent
\textbf{Evaluation Metrics.} 
To evaluate reconstruction performance, we adopt peak signal-to-noise ratio (PSNR) and structural similarity (SSIM) as quantitative evaluation metrics, following prior works~\cite{lin2023learning, lin2024c2rv}. Higher PSNR and SSIM values indicate superior image quality.

\subsection{Results}

\noindent
In the following experiments, the reconstruction model DiCE is first pretrained on the processed AbdomenAtlas-8K with HyViP pretraining and subsequently finetuned on various target datasets with different numbers of projection views.

\vspace{3pt}
\noindent
\textbf{Performance on various target datasets.} 
We compare the performance of DeepSparse (\ie, the finetuned model) against previous sparse-view reconstruction methods. Specifically, the comparison includes self-supervised methods (\ie, FDK~\cite{feldkamp1984practical}, SART~\cite{andersen1984simultaneous}, NAF~\cite{zha2022naf}, and NeRP~\cite{shen2022nerp}), data-driven denoising methods (\ie, FBPConvNet~\cite{jin2017deep}, FreeSeed~\cite{ma2023freeseed}, and BBDM~\cite{li2023bbdm}), and data-driven INR\footnote{INR: implicit neural representation}-based methods (\ie, PixelNeRF~\cite{yu2021pixelnerf}, DIF-Net~\cite{lin2023learning}, and C$^\text{2}$RV~\cite{lin2024c2rv}). We follow \cite{lin2024c2rv} to conduct experiments with different numbers of projection views (\ie, 6, 8, and 10) across two target datasets. The reconstruction resolution is set to $256^3$. \cgs{For self-supervised methods, their parameters are optimized per testing sample; for data-driven methods, their models are trained from scratch on the target set.} As shown in Table~\ref{tab:ssl} and \ref{tab:datadriven}, our DeepSparse significantly outperforms previous methods. Notably, compared to the previous state-of-the-art method C$^\text{2}$RV~\cite{lin2024c2rv}, our DeepSparse achieves \cgs{an improvement of 1$\sim$2 dB PSNR and 2$\sim$4\% SSIM on head/chest/abdomen datasets, and 3$\sim$4 dB PSNR and 3$\sim$8\% SSIM on pelvis and knee datasets.}

\begin{table}[t]
\centering
\caption{The number of model parameters (Param.) and reconstruction time of different methods. The reconstruction resolution is 256$^\text{3}$.}
\label{tab:efficiency}
\setlength{\tabcolsep}{8pt}
\resizebox{1.0\linewidth}{!}{
\begin{tabular}{l|c|c|c|c}
\toprule[1.2pt]
\multirow{2}{*}{Method} & \multirow{2}{*}{\begin{tabular}[c]{@{}c@{}}Param.\\ (M)\end{tabular}} & \multicolumn{3}{c}{Reconstruction Time (s)} \\ \cline{3-5} 
 & & 6-view & 8-view & 10-view \\ \hline \hline
FBPConvNet~\cite{jin2017deep} & 34.6 & 1.7 & 1.7 & 1.7 \\
FreeSeed~\cite{ma2023freeseed} & 8.7 & 3.7 & 3.7 & 3.7 \\
PixelNeRF~\cite{yu2021pixelnerf} & 24.7 & 40.4 & 57.6 & 71.2 \\
DIF-Net~\cite{lin2023learning} & 31.1 & 1.1 & 1.4 & 1.6 \\
C$^\text{2}$RV~\cite{lin2024c2rv} & 50.8 & 23.8 & 31.3 & 39.3 \\ \hline
DeepSparse (Ours) & 7.2 & 3.1 & 4.1 & 5.0 \\
\bottomrule[1.2pt]
\end{tabular}
}
\end{table}

\vspace{3pt}
\noindent
\textbf{Qualitative evaluation.} In Figures \ref{fig:vislung} and \ref{fig:visknee}, we visualize the ground-truth CT and CT reconstructed by different methods from only 6 projections. Compared to the previous state-of-the-art method, C$^\text{2}$RV~\cite{lin2024c2rv}, our DeepSparse reconstructs CT volumes with richer details, fewer artifacts, and clearer organ boundaries. These improvements have the potential to enhance the visualization of critical organs for accurate intraoperative navigation and facilitate the reconstruction of bone models for preoperative planning.

\vspace{3pt}
\noindent
\textbf{Efficiency Analysis.} In Table~\ref{tab:efficiency}, we compare the number of model parameters and the processing efficiency of different data-driven methods. Our DeepSparse reconstructs CT in just a few seconds, with a reconstruction speed 7.6$\times$ faster than C$^\text{2}$RV~\cite{lin2024c2rv}. Furthermore, compared to C$^\text{2}$RV~\cite{lin2024c2rv}, DeepSparse uses only $1/7$ of the model parameters (\ie, 7.2M \vs~50.8M) while achieving better reconstruction performance, with improvements of $\geq$1.0 dB in PSNR and $\geq$2.5\% in SSIM.

\subsection{Ablation Study}

\noindent
In this section, we conduct ablation studies on LUNA16~\cite{setio2017validation} dataset with reconstruction resolution of $256^3$ to 1.) compare different network designs of the reconstruction framework DiCE, 2.) analyze the effectiveness of the proposed three training stages, which include pretraining (HyViP) and two finetuning steps, 3.) investigate the robustness of the model in different data-insufficient scenarios, and \cgs{4.) explore the impact of different pretraining data scales.}

\begin{table}[t]
{
\centering
\caption{Ablation study on the number of scales $S$ and the volumetric resolution $r$. The model (DiCE) is evaluated on LUNA16~\cite{setio2017validation}, and PSNR (dB) is reported in the table.}
\label{tab:design}
\setlength{\tabcolsep}{13pt}
\resizebox{1.0\linewidth}{!}{
\begin{tabular}{c|c|l|l|l}
\toprule[1.2pt]
$S$ & $r$ & 6-View & 8-View & 10-View \\ \hline \hline
\red{\textbf{3}} & 32 & 28.95\tiny{\red{-0.74}} & 29.73\tiny{\red{-0.71}} & 30.54\tiny{\red{-0.53}} \\
\red{\textbf{5}} & 32 & 29.70\tiny{\green{+0.01}}  & 30.43\tiny{\red{-0.01}} & 31.07\tiny{\red{-0.00}} \\ \hline
4 & \red{\textbf{24}} & 29.41\tiny{\red{-0.28}} & 30.12\tiny{\red{-0.32}} & 30.89\tiny{\red{-0.18}} \\ 
4 & \red{\textbf{40}} & 29.71\tiny{\green{+0.02}} & 30.42\tiny{\red{-0.02}} & 31.08\tiny{\green{+0.01}} \\ \hline
4 & 32 & 29.69 & 30.44 & 31.07 \\
\bottomrule[1.2pt]
\end{tabular}
}
}
\end{table}

\vspace{3pt}
\noindent
\textbf{Network design of DiCE.} 
Table~\ref{tab:design} presents a comparison of different network designs for the reconstruction framework DiCE, focusing on the number of scales ($S$) and the volumetric resolution ($r$). While increasing the number of scales to 5 or raising the resolution to 40 slightly enhances reconstruction performance, the improvements are marginal and come at a significantly higher computational cost.

\vspace{3pt}
\noindent
\textbf{Pretraining \& two-step finetuning.}
In Table~\ref{tab:stages}, we compare the following training strategies: 1.) training the network on the target set from scratch; 2.) performing only the two-step finetuning on the network with randomly initialized parameters (without pretraining); and 3.) pretraining the network on a large-scale dataset followed by the second step of finetuning. The results demonstrate that both pretraining and the denoising processes contribute to performance improvements, and combining them yields the best results. \cgs{Notably, even with pretraining and finetuning step-2, the performance of the model is slightly lower than training from scratch. The reason is that finetuning with only step-2 requires the model to address several gaps (\eg, image style, the number of views to generate 2D/3D features) simultaneously while the encoders remain fixed. This constraint significantly hinders convergence, resulting in suboptimal performance.}

\begin{table}[t]
{
\centering
\caption{Ablation study on three training stages, including the pretraining and 2-step finetuning. The model is evaluated on LUNA16~\cite{setio2017validation}. PSNR (dB) is reported in the table.}
\label{tab:stages}
\setlength{\tabcolsep}{3pt}
\resizebox{1.0\linewidth}{!}{
\begin{tabular}{c|c|cc|c|c|c}
\toprule[1.2pt]
\multirow{2}{*}{\cgs{From Scratch}} & \multirow{2}{*}{Pretrain} & \multicolumn{2}{c|}{Finetune} & \multirow{2}{*}{6-View} & \multirow{2}{*}{8-View} & \multirow{2}{*}{10-View} \\ \cline{3-4}
 & & Step-1 & Step-2 & & & \\ \hline \hline
 \cgs{\checkmark} & & & & 29.69 & 30.44 & 31.07 \\
 & & \checkmark & \checkmark & 29.73 & 30.55 & 31.09 \\
 & \checkmark & & \checkmark & 29.16 & 29.84 & 30.36 \\ \hline
 & \checkmark & \checkmark & \checkmark & 30.22 & 31.14 & 31.86 \\
\bottomrule[1.2pt]
\end{tabular}
}
}
\end{table}

\begin{table}[t]
{
\centering
\caption{Ablation study on the robustness in different data-insufficient scenarios. The model is evaluated on 6-view LUNA16~\cite{setio2017validation}. PSNR (dB) and SSIM ($\times$10$^\text{-2}$) are reported in the table. w/o pretraining: to train the model (DiCE) from scratch. w/ pretraining: to pretrain the model then conduct 2-step finetuning.}
\label{tab:fewshot}
\setlength{\tabcolsep}{13pt}
\resizebox{1.0\linewidth}{!}{
\begin{tabular}{l|c|c}
\toprule[1.2pt]
\# Data & w/o Pretraining & w/ Pretraining \\ \hline\hline
100\% (738) & 29.69$|$88.68 & 30.22$|$89.96 \\ \hline
50\% (369)  & 29.61$|$88.57 & 30.21$|$89.93 \\
20\% (147)  & 28.43$|$87.23 & 29.70$|$88.54 \\
10\% (73)   & 27.13$|$86.58 & 28.35$|$87.85 \\
\bottomrule[1.2pt]
\end{tabular}
}
}
\end{table}

\vspace{3pt}
\noindent
\textbf{Robustness in data-insufficient scenarios.}
We evaluate the robustness of the pretrained model on various data-insufficient target sets. The results in Table~\ref{tab:fewshot} show that with pretraining, finetuning using only 20\% of the target data achieves reconstruction performance comparable to training from scratch on the full (100\%) target set.

\vspace{3pt}
\noindent
\cgs{\textbf{Data Scale of Pretraining.} To assess the scalability of our approach, we conducted a study examining the influence of pretraining dataset size. We trained the model using subsets of the AbdomenAtlas-8K dataset, ranging from 1,000 to the full 8,407 cases. As shown in Table~\ref{tab:datascale}, we observe a consistent performance improvement across all sparse-view settings (\ie, 6, 8, and 10 views) as the scale of pretraining data increases.} 

\begin{table}[t]
{
\centering
\caption{\cgs{Ablation study on pretraining data scale. The model is evaluated on LUNA16~\cite{setio2017validation}. PSNR (dB) and SSIM ($\times10^{-2}$)) are reported in the table.}}
\label{tab:datascale}
\setlength{\tabcolsep}{5pt}
\resizebox{1.0\linewidth}{!}{
\begin{tabular}{c|c|c|c}
\toprule[1.2pt]
\# Pretrain Data & 6-View & 8-View & 10-View \\ \hline \hline
8,407 & 30.22$|$89.96 & 31.14$|$90.76 & 31.86$|$91.41 \\
4,000 & 30.11$|$89.72 & 31.02$|$90.51 & 31.68$|$91.15 \\
2,000 & 29.85$|$89.03 & 30.83$|$90.08 & 31.49$|$90.52 \\ 
1,000 & 29.71$|$88.81 & 30.66$|$89.70 & 31.28$|$90.24 \\
0: train from scratch & 29.69$|$88.68 & 30.44$|$89.55 & 31.07$|$90.25 \\
\bottomrule[1.2pt]
\end{tabular}
}
\label{tab:datascale}
}
\end{table}
\begin{table}[t]
{
\centering
\caption{\cgs{VIF scores of data-driven methods on chest (LUNA16~\cite{setio2017validation}) and knee (Lin et al. ~\cite{lin2023learning}) datasets with various numbers of projection views. Higher VIF indicate better performance.}}
\label{tab:vif_comparison}
\setlength{\tabcolsep}{3pt}
\resizebox{1.0\linewidth}{!}{
\begin{tabular}{l|c|c|c|c|c|c}
\toprule[1.2pt]
 & \multicolumn{3}{c|}{LUNA} & \multicolumn{3}{c}{Knee} \\
\hline
Method & 6-View & 8-View & 10-View & 6-View & 8-View & 10-View \\ \hline \hline
FBPConvNet~\cite{jin2017deep} & 0.1609 & 0.1765 & 0.2010 & 0.2777 & 0.3069 & 0.3280 \\
Freeseed~\cite{ma2023freeseed} & 0.1513 & 0.1728 & 0.1931 & 0.2634 & 0.2698 & 0.3071 \\
BBDM~\cite{li2023bbdm} & 0.1426 & 0.1680 & 0.2000 & 0.2712 & 0.2990 & 0.3168 \\
PixelNeRF~\cite{yu2021pixelnerf} & 0.1404 & 0.1627 & 0.1858 & 0.1695 & 0.1913 & 0.2047 \\
DIF-Net~\cite{lin2023learning} & 0.1489 & 0.1229 & 0.1442 & 0.2143 & 0.2306 & 0.2445 \\
C$^\text{2}$RV~\cite{lin2024c2rv} & 0.2361 & 0.2523 & 0.2627 & 0.2849 & 0.3147 & 0.3314 \\
Ours & \textbf{0.2625} & \textbf{0.2801} & \textbf{0.3012} & \textbf{0.3817} & \textbf{0.4138} & \textbf{0.4391} \\
\bottomrule[1.2pt]
\end{tabular}
}
\label{tab:vif}
}
\end{table}

\section{Discussion}

\noindent
\cgs{
In this section, we extend the evaluation of our proposed DeepSparse beyond standard quantitative and qualitative metrics. Specifically, we 1.) assess reconstruction quality using perceptually relevant metrics and discuss the feasiblity of evaluation with real measurements; 2.) analyze the clinical utility of our reconstructed CTs in specific downstream applications; and 3.) examine the model's performance under non-ideal imaging conditions.
}

\subsection{\cgs{A Metrics}}

\noindent
\cgs{
\textbf{Perceptual Quality Assessment:} Metrics such as PSNR and SSIM may not correlate well with clinical diagnostic quality. We follow the previous study \cite{liu2024imaging} to expand the evaluation to include Visual Information Fidelity (VIF), as VIF exhibits a higher correlation with human reader ratings for CT images compared to pixel-based metrics~\cite{liu2024imaging}. Results in Table~\ref{tab:vif} highlights that our proposed method consistently outperforms previous methods in VIF scores. This quantitative advantage corroborates the qualitative observations (Figures~\ref{fig:vislung} and \ref{fig:visknee}), indicating that our method recovers perceptually relevant visual information more effectively than other approaches.
}

\vspace{3pt}
\noindent
\cgs{
\textbf{Evaluation on Real-World Measurements:} The transition from simulated to real-world acquisition can be regard as a specific type of domain shift. Our experiments demonstrates that our model effectively bridges domain gaps arising from varying anatomical structures, patient cohorts, and imaging protocols. The model’s strong robustness suggests that the model should have the potential generalizability to adapt to real-world projection data. However, experimental validation on real-world CBCT scanner is currently infeasible due to the lack of access to raw projection data from commercial scanning devices. Consequently, this study relies on high-fidelity simulations to approximate clinical conditions, while future work will prioritize vendor collaboration to facilitate validation on real measured projections.
}

\begin{figure}[t]
    \centering
    \includegraphics[width=1.0\linewidth]{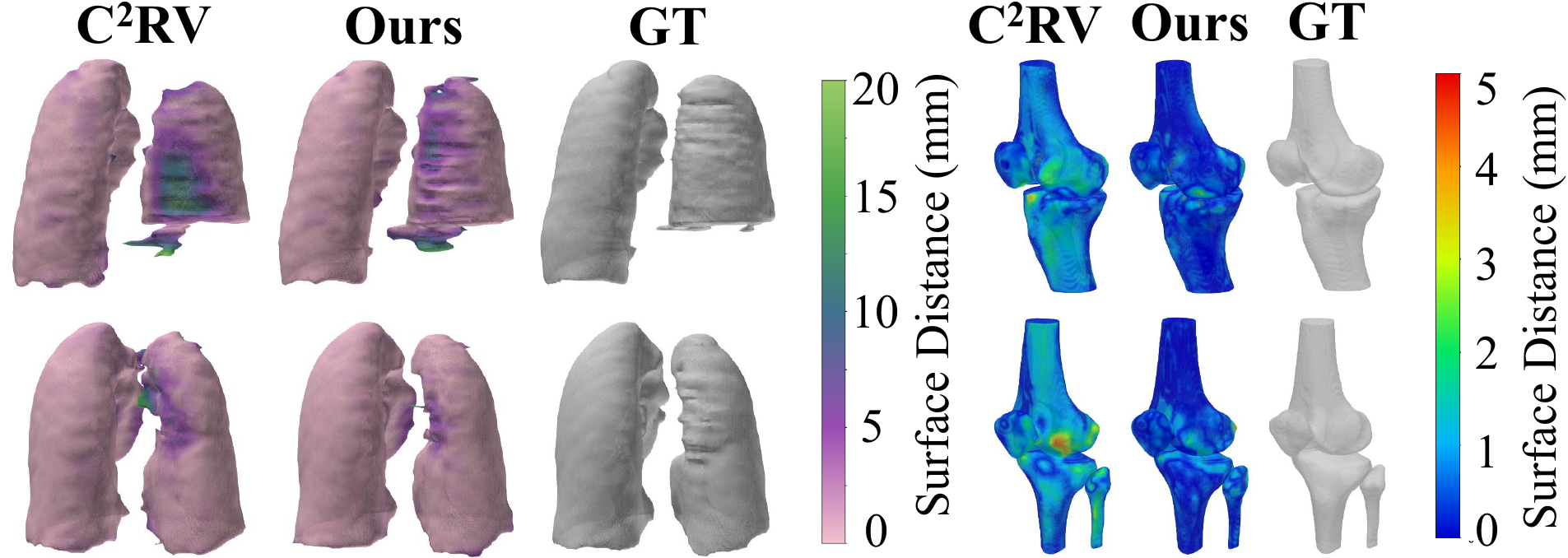}
    \vspace{-0.5cm}
    \caption{\cgs{Visualization of surface errors for lung and knee bone structures, comparing automatic segmentations from 6-view reconstructed CTs against those from the ground truth CTs.}}
    \label{fig:segmentation}
\end{figure}

\begin{table}[t]
{
\centering
\caption{\cgs{Segmentation performance of lung and knee bone structures on 6-view reconstructed CTs. Evaluation was performed based on masks segmented from reconstructed CTs and those from ground-truth CTs using automated segmentation tools. Metrics including Dice(\%), ASSD (mm), and HD95(mm) are reported in the table.}}
\label{tab:segmentation_metrics}
\setlength{\tabcolsep}{3pt}
\resizebox{1.0\linewidth}{!}{
\begin{tabular}{l|c|c|c|c|c|c}
\toprule[1.2pt]
 & \multicolumn{3}{c|}{LUNA16~\cite{setio2017validation} (Lung)} & \multicolumn{3}{c}{Lin\etal~\cite{lin2023learning} (Knee)} \\
\hline
Method & Dice $\uparrow$ & ASSD $\downarrow$ & HD95 $\downarrow$ & Dice $\uparrow$ & ASSD $\downarrow$ & HD95 $\downarrow$ \\ \hline \hline
C$^\text{2}$RV & 96.91 & 0.3936 & 1.637 & 96.41 & 0.3492 & 1.268 \\
Ours & 97.02 & 0.3926 & 1.572 & 96.82 & 0.2798 & 1.125 \\
\bottomrule[1.2pt]
\end{tabular}
}
\label{tab:segmentation}
}
\end{table}
\begin{figure}[t]
    \centering
    \includegraphics[width=1.0\linewidth]{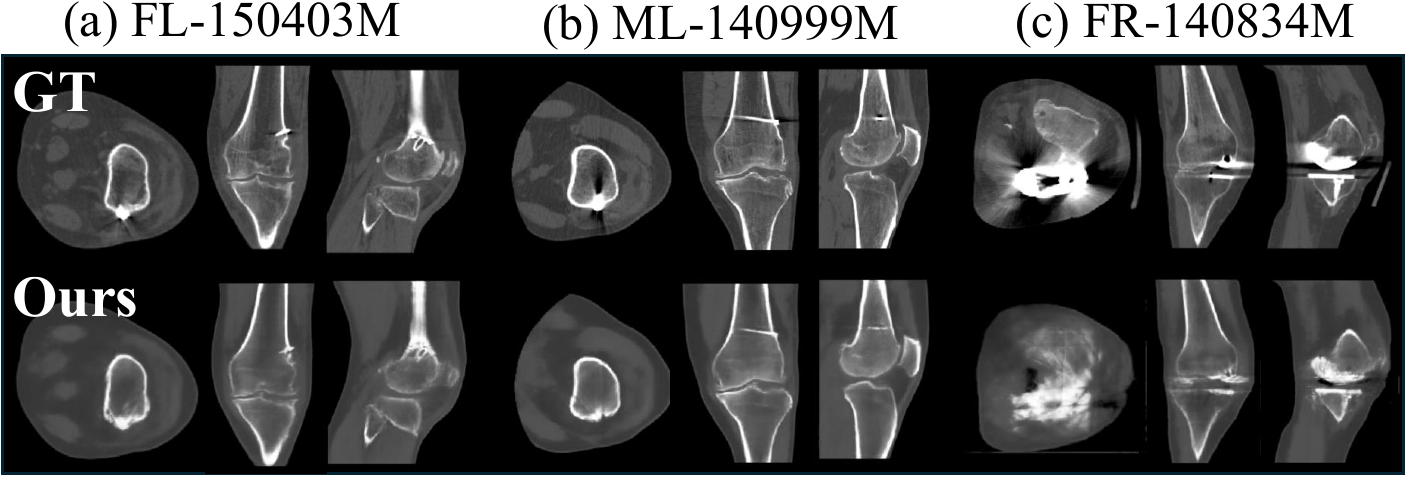}
    \vspace{-0.6cm}
    \caption{\cgs{Cases with metal implants. Reconstruction with 6 projections on the knee dataset.}}
    \label{fig:metal_implant}
\end{figure}

\subsection{\cgs{Clinical Relevance}}
\noindent
\cgs{
To validate the clinical utility of the proposed method, we assessed whether the reconstructed images preserve sufficient structural fidelity to support downstream clinical tasks, such as organ volumetry and preoperative planning. To achieve this, we performed automated segmentation on sparse-view reconstructed CTs and ground-truth CTs, and compare their differences using Dice score, Average Symmetric Surface Distance (ASSD), and $95\%$ Hausdorff Distance (HD95).
For LUNA16~\cite{setio2017validation}, we employed LungMask\footnote{https://github.com/JoHof/lungmask}~\cite{hofmanninger2020automatic}, which is specifically optimized for automated lung segmentation in the presence of diverse pathologies and visual variability, making it a robust evaluator for our reconstruction quality. For the knee dataset (Lin~\etal)~\cite{lin2023learning}, we trained a segmentation model using nnUNet framework\footnote{https://github.com/MIC-DKFZ/nnUNet}~\cite{isensee2021nnu}, which provides automatic architecture and training hyperparameters adaptation ensure a strong segmentation baseline.
}

\cgs{
As presented in Figure~\ref{fig:segmentation} and Table~\ref{tab:segmentation}, the reconstruction of key anatomical structures remains accurate even under sparse-view inputs, and our method consistently outperforms C$^\text{2}$RV~\cite{lin2024c2rv} in both segmentation tasks. These results highlight the potential of the reconstructed CT in several critical clinical scenarios, including pulmonary volumetry, preoperative planning for procedures such as high tibial osteotomy (HTO), and providing geometric references for intraoperative navigation. More importantly, our DeepSparse serves as a foundational framework, providing the robust generalization necessary for diverse clinical deployments.
}

\subsection{\cgs{Non-Ideal Scenarios -- Metal Implants}}

\noindent
\cgs{
The projection acquisition process is not always ideal. Metal artifacts represent a significant challenge in CT reconstruction due to beam hardening and photon starvation. To validate the robustness of our model in non-ideal scenarios, we collect three samples with metal implants from the same hospital where the knee dataset (Lin~\etal~\cite{lin2023learning}) is collected. We finetuned the model on the original implant-free knee dataset, and then test it on data with metal implants. Reconstructed CT are visualized in Figure~\ref{fig:metal_implant}, showing that our model performs fairly on cases with minor bone disruption while worse on cases with major disruption.}

\cgs{
Specifically, for the cases (FL-150403M and ML-140999M) involving small implants (e.g., surgical staples and screws) that cause minimal disruption to native bone structure, our model demonstrates good robustness, successfully reconstructing the bone anatomy with high quality. 
However, for the case (FR-140834M) involving large implants (e.g., total joint replacements) that significantly replace or alter the bone topology, our model currently struggles to accurately reconstruct the bone anatomy. This is expected, as the unique geometric features of these large artificial components were not represented in the training distribution.
}

\cgs{
These results highlight a clear path for future improvement. To enhance robustness for large implants, the training dataset must be expanded to include diverse implant examples. Furthermore, since clinical applications often require separating the implant from human anatomy, and implant shapes are highly standardized, a promising direction is to incorporate parameterized templates as geometric priors to model these distinct components explicitly.
}

\section{Conclusion}

\noindent
In this work, we introduce DeepSparse, the first foundation model for sparse-view CBCT reconstruction. Specifically, we propose a novel and effective reconstruction network, DiCE, which simplifies the 2D feature extraction and utilizes a 3D decoder to efficiently aggregate multi-scale features, enhancing the 3D representation. To improve generalizability and robustness, we pretrain the network on a large-scale dataset with hybrid view sampling. Furthermore, we introduce a two-step finetuning process to effectively adapt the pretrained model to various target datasets. Experiments and ablation studies demonstrate that DeepSparse achieves superior reconstruction performance compared to previous state-of-the-art methods. In the future, we aim to 1.) simplify the finetuning process and develop a more generalized reconstruction model, extending its applicability to a broader range of scenarios, \cgs{2.) collobrate with scanner vendor to conduct validation on real measurements, and 3.) improve the robustness of the reconstruction model in non-ideal scanning scenarios.}

\bibliographystyle{IEEEtran}
\bibliography{tmi}

\begin{thebibliography}{10}
\providecommand{\url}[1]{#1}
\csname url@samestyle\endcsname
\providecommand{\newblock}{\relax}
\providecommand{\bibinfo}[2]{#2}
\providecommand{\BIBentrySTDinterwordspacing}{\spaceskip=0pt\relax}
\providecommand{\BIBentryALTinterwordstretchfactor}{4}
\providecommand{\BIBentryALTinterwordspacing}{\spaceskip=\fontdimen2\font plus
\BIBentryALTinterwordstretchfactor\fontdimen3\font minus \fontdimen4\font\relax}
\providecommand{\BIBforeignlanguage}[2]{{%
\expandafter\ifx\csname l@#1\endcsname\relax
\typeout{** WARNING: IEEEtran.bst: No hyphenation pattern has been}%
\typeout{** loaded for the language `#1'. Using the pattern for}%
\typeout{** the default language instead.}%
\else
\language=\csname l@#1\endcsname
\fi
#2}}
\providecommand{\BIBdecl}{\relax}
\BIBdecl

\bibitem{scarfe2006clinical}
W.~C. Scarfe, A.~G. Farman, P.~Sukovic \emph{et~al.}, ``Clinical applications of cone-beam computed tomography in dental practice,'' \emph{Journal-Canadian Dental Association}, vol.~72, no.~1, p.~75, 2006.

\bibitem{brenner2007computed}
D.~J. Brenner and E.~J. Hall, ``Computed tomography—an increasing source of radiation exposure,'' \emph{New England journal of medicine}, vol. 357, no.~22, pp. 2277--2284, 2007.

\bibitem{miglioretti2013use}
D.~L. Miglioretti, E.~Johnson, A.~Williams, R.~T. Greenlee, S.~Weinmann, L.~I. Solberg, H.~S. Feigelson, D.~Roblin, M.~J. Flynn, N.~Vanneman \emph{et~al.}, ``The use of computed tomography in pediatrics and the associated radiation exposure and estimated cancer risk,'' \emph{JAMA pediatrics}, vol. 167, no.~8, pp. 700--707, 2013.

\bibitem{pearce2012radiation}
M.~S. Pearce, J.~A. Salotti, M.~P. Little, K.~McHugh, C.~Lee, K.~P. Kim, N.~L. Howe, C.~M. Ronckers, P.~Rajaraman, A.~W. Craft \emph{et~al.}, ``Radiation exposure from ct scans in childhood and subsequent risk of leukaemia and brain tumours: a retrospective cohort study,'' \emph{The Lancet}, vol. 380, no. 9840, pp. 499--505, 2012.

\bibitem{lee2004diagnostic}
C.~I. Lee, A.~H. Haims, E.~P. Monico, J.~A. Brink, and H.~P. Forman, ``Diagnostic ct scans: assessment of patient, physician, and radiologist awareness of radiation dose and possible risks,'' \emph{Radiology}, vol. 231, no.~2, pp. 393--398, 2004.

\bibitem{jin2017deep}
K.~H. Jin, M.~T. McCann, E.~Froustey, and M.~Unser, ``Deep convolutional neural network for inverse problems in imaging,'' \emph{IEEE Transactions on Image Processing}, vol.~26, no.~9, pp. 4509--4522, 2017.

\bibitem{han2016deep}
Y.~S. Han, J.~Yoo, and J.~C. Ye, ``Deep residual learning for compressed sensing ct reconstruction via persistent homology analysis,'' \emph{arXiv preprint arXiv:1611.06391}, 2016.

\bibitem{guo2023self}
J.~Guo, X.~Xu, and H.~Zhao, ``Self-supervised learning for enhancing geometrical modeling in 3d-aware generative adversarial network,'' \emph{arXiv preprint arXiv:2312.11856}, 2023.

\bibitem{zhang2018sparse}
Z.~Zhang, X.~Liang, X.~Dong, Y.~Xie, and G.~Cao, ``A sparse-view ct reconstruction method based on combination of densenet and deconvolution,'' \emph{IEEE transactions on medical imaging}, vol.~37, no.~6, pp. 1407--1417, 2018.

\bibitem{wang2018conditional}
J.~Wang, Y.~Zhao, J.~H. Noble, and B.~M. Dawant, ``Conditional generative adversarial networks for metal artifact reduction in ct images of the ear,'' in \emph{Medical Image Computing and Computer Assisted Intervention--MICCAI 2018: 21st International Conference, Granada, Spain, September 16-20, 2018, Proceedings, Part I}.\hskip 1em plus 0.5em minus 0.4em\relax Springer, 2018, pp. 3--11.

\bibitem{huang2018metal}
X.~Huang, J.~Wang, F.~Tang, T.~Zhong, and Y.~Zhang, ``Metal artifact reduction on cervical ct images by deep residual learning,'' \emph{Biomedical engineering online}, vol.~17, pp. 1--15, 2018.

\bibitem{ma2023freeseed}
C.~Ma, Z.~Li, J.~Zhang, Y.~Zhang, and H.~Shan, ``Freeseed: Frequency-band-aware and self-guided network for sparse-view ct reconstruction,'' in \emph{International Conference on Medical Image Computing and Computer-Assisted Intervention}.\hskip 1em plus 0.5em minus 0.4em\relax Springer, 2023, pp. 250--259.

\bibitem{wu2021drone}
W.~Wu, D.~Hu, C.~Niu, H.~Yu, V.~Vardhanabhuti, and G.~Wang, ``Drone: Dual-domain residual-based optimization network for sparse-view ct reconstruction,'' \emph{IEEE Transactions on Medical Imaging}, vol.~40, no.~11, pp. 3002--3014, 2021.

\bibitem{he2020radon}
J.~He, Y.~Wang, and J.~Ma, ``Radon inversion via deep learning,'' \emph{IEEE transactions on medical imaging}, vol.~39, no.~6, pp. 2076--2087, 2020.

\bibitem{song2021solving}
Y.~Song, L.~Shen, L.~Xing, and S.~Ermon, ``Solving inverse problems in medical imaging with score-based generative models,'' \emph{arXiv preprint arXiv:2111.08005}, 2021.

\bibitem{chung2023solving}
H.~Chung, D.~Ryu, M.~T. McCann, M.~L. Klasky, and J.~C. Ye, ``Solving 3d inverse problems using pre-trained 2d diffusion models,'' in \emph{Proceedings of the IEEE/CVF Conference on Computer Vision and Pattern Recognition}, 2023, pp. 22\,542--22\,551.

\bibitem{lin2019dudonet}
W.-A. Lin, H.~Liao, C.~Peng, X.~Sun, J.~Zhang, J.~Luo, R.~Chellappa, and S.~K. Zhou, ``Dudonet: Dual domain network for ct metal artifact reduction,'' in \emph{Proceedings of the IEEE/CVF Conference on Computer Vision and Pattern Recognition}, 2019, pp. 10\,512--10\,521.

\bibitem{wang2021dudotrans}
C.~Wang, K.~Shang, H.~Zhang, Q.~Li, Y.~Hui, and S.~K. Zhou, ``Dudotrans: dual-domain transformer provides more attention for sinogram restoration in sparse-view ct reconstruction,'' \emph{arXiv preprint arXiv:2111.10790}, 2021.

\bibitem{lin2023learning}
Y.~Lin, Z.~Luo, W.~Zhao, and X.~Li, ``Learning deep intensity field for extremely sparse-view cbct reconstruction,'' in \emph{Medical Image Computing and Computer Assisted Intervention -- MICCAI 2023}.\hskip 1em plus 0.5em minus 0.4em\relax Cham: Springer Nature Switzerland, 2023, pp. 13--23.

\bibitem{mildenhall2021nerf}
B.~Mildenhall, P.~P. Srinivasan, M.~Tancik, J.~T. Barron, R.~Ramamoorthi, and R.~Ng, ``Nerf: Representing scenes as neural radiance fields for view synthesis,'' \emph{Communications of the ACM}, vol.~65, no.~1, pp. 99--106, 2021.

\bibitem{zha2022naf}
R.~Zha, Y.~Zhang, and H.~Li, ``Naf: Neural attenuation fields for sparse-view cbct reconstruction,'' in \emph{Medical Image Computing and Computer Assisted Intervention--MICCAI 2022: 25th International Conference, Singapore, September 18--22, 2022, Proceedings, Part VI}.\hskip 1em plus 0.5em minus 0.4em\relax Springer, 2022, pp. 442--452.

\bibitem{lin2024learning}
Y.~Lin, H.~Wang, J.~Chen, and X.~Li, ``Learning 3d gaussians for extremely sparse-view cone-beam ct reconstruction,'' in \emph{International Conference on Medical Image Computing and Computer-Assisted Intervention}.\hskip 1em plus 0.5em minus 0.4em\relax Springer, 2024, pp. 425--435.

\bibitem{lin2024c2rv}
Y.~Lin, J.~Yang, H.~Wang, X.~Ding, W.~Zhao, and X.~Li, ``C{\textasciicircum}2rv: Cross-regional and cross-view learning for sparse-view cbct reconstruction,'' in \emph{Proceedings of the IEEE/CVF Conference on Computer Vision and Pattern Recognition (CVPR)}, June 2024, pp. 11\,205--11\,214.

\bibitem{fang2022snaf}
Y.~Fang, L.~Mei, C.~Li, Y.~Liu, W.~Wang, Z.~Cui, and D.~Shen, ``Snaf: Sparse-view cbct reconstruction with neural attenuation fields,'' \emph{arXiv preprint arXiv:2211.17048}, 2022.

\bibitem{shen2022nerp}
L.~Shen, J.~Pauly, and L.~Xing, ``Nerp: implicit neural representation learning with prior embedding for sparsely sampled image reconstruction,'' \emph{IEEE Transactions on Neural Networks and Learning Systems}, 2022.

\bibitem{zha2024r}
R.~Zha, T.~J. Lin, Y.~Cai, J.~Cao, Y.~Zhang, and H.~Li, ``R2-gaussian: Rectifying radiative gaussian splatting for tomographic reconstruction,'' \emph{arXiv preprint arXiv:2405.20693}, 2024.

\bibitem{ronneberger2015u}
O.~Ronneberger, P.~Fischer, and T.~Brox, ``U-net: Convolutional networks for biomedical image segmentation,'' in \emph{Medical Image Computing and Computer-Assisted Intervention--MICCAI 2015: 18th International Conference, Munich, Germany, October 5-9, 2015, Proceedings, Part III 18}.\hskip 1em plus 0.5em minus 0.4em\relax Springer, 2015, pp. 234--241.

\bibitem{huang2017densely}
G.~Huang, Z.~Liu, L.~Van Der~Maaten, and K.~Q. Weinberger, ``Densely connected convolutional networks,'' in \emph{Proceedings of the IEEE conference on computer vision and pattern recognition}, 2017, pp. 4700--4708.

\bibitem{feldkamp1984practical}
L.~A. Feldkamp, L.~C. Davis, and J.~W. Kress, ``Practical cone-beam algorithm,'' \emph{Josa a}, vol.~1, no.~6, pp. 612--619, 1984.

\bibitem{gordon1970algebraic}
R.~Gordon, R.~Bender, and G.~T. Herman, ``Algebraic reconstruction techniques (art) for three-dimensional electron microscopy and x-ray photography,'' \emph{Journal of theoretical Biology}, vol.~29, no.~3, pp. 471--481, 1970.

\bibitem{andersen1984simultaneous}
A.~H. Andersen and A.~C. Kak, ``Simultaneous algebraic reconstruction technique (sart): a superior implementation of the art algorithm,'' \emph{Ultrasonic imaging}, vol.~6, no.~1, pp. 81--94, 1984.

\bibitem{pan2006variable}
J.~Pan, T.~Zhou, Y.~Han, and M.~Jiang, ``Variable weighted ordered subset image reconstruction algorithm,'' \emph{International Journal of Biomedical Imaging}, vol. 2006, 2006.

\bibitem{jiang2022mfct}
Y.~Jiang, ``Mfct-gan: multi-information network to reconstruct ct volumes for security screening,'' \emph{Journal of Intelligent Manufacturing and Special Equipment}, 2022.

\bibitem{shen2019patient}
L.~Shen, W.~Zhao, and L.~Xing, ``Patient-specific reconstruction of volumetric computed tomography images from a single projection view via deep learning,'' \emph{Nature biomedical engineering}, vol.~3, no.~11, pp. 880--888, 2019.

\bibitem{ying2019x2ct}
X.~Ying, H.~Guo, K.~Ma, J.~Wu, Z.~Weng, and Y.~Zheng, ``X2ct-gan: reconstructing ct from biplanar x-rays with generative adversarial networks,'' in \emph{Proceedings of the IEEE/CVF conference on computer vision and pattern recognition}, 2019, pp. 10\,619--10\,628.

\bibitem{kyung2023perspective}
D.~Kyung, K.~Jo, J.~Choo, J.~Lee, and E.~Choi, ``Perspective projection-based 3d ct reconstruction from biplanar x-rays,'' in \emph{ICASSP 2023-2023 IEEE International Conference on Acoustics, Speech and Signal Processing (ICASSP)}.\hskip 1em plus 0.5em minus 0.4em\relax IEEE, 2023, pp. 1--5.

\bibitem{ruckert2022neat}
D.~R{\"u}ckert, Y.~Wang, R.~Li, R.~Idoughi, and W.~Heidrich, ``Neat: Neural adaptive tomography,'' \emph{ACM Transactions on Graphics (TOG)}, vol.~41, no.~4, pp. 1--13, 2022.

\bibitem{alkaeed2025open}
M.~Alkaeed, S.~Abioye, A.~Qayyum, Y.~M. Mekki, I.~Berrou, M.~Abdallah, A.~Al-Fuqaha, M.~Bilal, and J.~Qadir, ``Open foundation models in healthcare: Challenges, paradoxes, and opportunities with genai driven personalized prescription,'' \emph{arXiv preprint arXiv:2502.04356}, 2025.

\bibitem{wang2025triad}
S.~Wang, M.~Safari, Q.~Li, C.-W. Chang, R.~L. Qiu, J.~Roper, D.~S. Yu, and X.~Yang, ``Triad: Vision foundation model for 3d magnetic resonance imaging,'' \emph{arXiv preprint arXiv:2502.14064}, 2025.

\bibitem{blankemeier2024merlin}
L.~Blankemeier, J.~P. Cohen, A.~Kumar, D.~Van~Veen, S.~J.~S. Gardezi, M.~Paschali, Z.~Chen, J.-B. Delbrouck, E.~Reis, C.~Truyts \emph{et~al.}, ``Merlin: A vision language foundation model for 3d computed tomography,'' \emph{Research Square}, pp. rs--3, 2024.

\bibitem{yao2024eva}
J.~Yao, X.~Wang, Y.~Song, H.~Zhao, J.~Ma, Y.~Chen, W.~Liu, and B.~Wang, ``Eva-x: A foundation model for general chest x-ray analysis with self-supervised learning,'' \emph{arXiv preprint arXiv:2405.05237}, 2024.

\bibitem{chen2024chexagent}
Z.~Chen, M.~Varma, J.-B. Delbrouck, M.~Paschali, L.~Blankemeier, D.~Van~Veen, J.~M.~J. Valanarasu, A.~Youssef, J.~P. Cohen, E.~P. Reis \emph{et~al.}, ``Chexagent: Towards a foundation model for chest x-ray interpretation,'' \emph{arXiv preprint arXiv:2401.12208}, 2024.

\bibitem{li2025towards}
C.-Y. Li, K.-J. Chang, C.-F. Yang, H.-Y. Wu, W.~Chen, H.~Bansal, L.~Chen, Y.-P. Yang, Y.-C. Chen, S.-P. Chen \emph{et~al.}, ``Towards a holistic framework for multimodal llm in 3d brain ct radiology report generation,'' \emph{Nature Communications}, vol.~16, no.~1, p. 2258, 2025.

\bibitem{huang2024enhancing}
W.~Huang, C.~Li, H.-Y. Zhou, H.~Yang, J.~Liu, Y.~Liang, H.~Zheng, S.~Zhang, and S.~Wang, ``Enhancing representation in radiography-reports foundation model: A granular alignment algorithm using masked contrastive learning,'' \emph{Nature Communications}, vol.~15, no.~1, p. 7620, 2024.

\bibitem{terris2025reconstruct}
M.~Terris, S.~Hurault, M.~Song, and J.~Tachella, ``Reconstruct anything model: a lightweight foundation model for computational imaging,'' \emph{arXiv preprint arXiv:2503.08915}, 2025.

\bibitem{liu2024imaging}
Y.~Liu, R.~Ge, Y.~He, Z.~Wu, S.~Yang, Y.~Gao, C.~You, G.~Wang, Y.~Chen, and S.~Li, ``Imaging foundation model for universal enhancement of non-ideal measurement ct,'' \emph{arXiv preprint arXiv:2410.01591}, 2024.

\bibitem{liu2025foundation}
Y.~Liu, C.~Li, H.~Liu, C.~Yang, and Y.~Yuan, ``Foundation model-guided gaussian splatting for 4d reconstruction of deformable tissues,'' \emph{IEEE Transactions on Medical Imaging}, 2025.

\bibitem{razavi2019generating}
A.~Razavi, A.~Van~den Oord, and O.~Vinyals, ``Generating diverse high-fidelity images with vq-vae-2,'' \emph{Advances in neural information processing systems}, vol.~32, 2019.

\bibitem{qu2023abdomenatlas}
C.~Qu, T.~Zhang, H.~Qiao, Y.~Tang, A.~L. Yuille, Z.~Zhou \emph{et~al.}, ``Abdomenatlas-8k: Annotating 8,000 ct volumes for multi-organ segmentation in three weeks,'' \emph{Advances in Neural Information Processing Systems}, vol.~36, pp. 36\,620--36\,636, 2023.

\bibitem{setio2017validation}
A.~A.~A. Setio, A.~Traverso, T.~De~Bel, M.~S. Berens, C.~Van Den~Bogaard, P.~Cerello, H.~Chen, Q.~Dou, M.~E. Fantacci, B.~Geurts \emph{et~al.}, ``Validation, comparison, and combination of algorithms for automatic detection of pulmonary nodules in computed tomography images: the luna16 challenge,'' \emph{Medical image analysis}, vol.~42, pp. 1--13, 2017.

\bibitem{imambi2021pytorch}
S.~Imambi, K.~B. Prakash, and G.~Kanagachidambaresan, ``Pytorch,'' \emph{Programming with TensorFlow: solution for edge computing applications}, pp. 87--104, 2021.

\bibitem{cipriano2022deep}
``Deep segmentation of the mandibular canal: a new 3d annotated dataset of cbct volumes,'' \emph{IEEE Access}, 2022.

\bibitem{alves2024panorama}
N.~Alves, M.~Schuurmans, D.~Rutkowski, D.~Yakar, I.~Haldorsen, M.~Liedenbaum, A.~Molven, P.~Vendittelli, G.~Litjens, J.~Hermans, and H.~Huisman, ``The panorama study protocol: Pancreatic cancer diagnosis - radiologists meet ai,'' \emph{Zenodo}, 2024.

\bibitem{liu2025automatic}
Y.~Liu, S.~Yibulayimu, G.~Zhu, C.~Shi, C.~Liang, C.~Zhao, X.~Wu, Y.~Sang, and Y.~Wang, ``Automatic pelvic fracture segmentation: a deep learning approach and benchmark dataset,'' \emph{Frontiers in Medicine}, 2025.

\bibitem{li2023bbdm}
B.~Li, K.~Xue, B.~Liu, and Y.-K. Lai, ``Bbdm: Image-to-image translation with brownian bridge diffusion models,'' in \emph{Proceedings of the IEEE/CVF Conference on Computer Vision and Pattern Recognition}, 2023, pp. 1952--1961.

\bibitem{yu2021pixelnerf}
A.~Yu, V.~Ye, M.~Tancik, and A.~Kanazawa, ``pixelnerf: Neural radiance fields from one or few images,'' in \emph{Proceedings of the IEEE/CVF Conference on Computer Vision and Pattern Recognition}, 2021, pp. 4578--4587.

\bibitem{hofmanninger2020automatic}
J.~Hofmanninger, F.~Prayer, J.~Pan, S.~R{\"o}hrich, H.~Prosch, and G.~Langs, ``Automatic lung segmentation in routine imaging is primarily a data diversity problem, not a methodology problem,'' \emph{European Radiology Experimental}, vol.~4, no.~1, pp. 1--13, 2020.

\bibitem{isensee2021nnu}
F.~Isensee, P.~F. Jaeger, S.~A. Kohl, J.~Petersen, and K.~H. Maier-Hein, ``nnu-net: a self-configuring method for deep learning-based biomedical image segmentation,'' \emph{Nature methods}, vol.~18, no.~2, pp. 203--211, 2021.

\end{thebibliography}

\end{document}